\documentclass[singlecolumn,english,aps,prl,showpacs,floatfix,superscriptaddress,notitlepage]{revtex4-1}

\usepackage[T1]{fontenc}
\usepackage[latin9]{inputenc}
\usepackage{amsmath}
\usepackage{booktabs}
\usepackage{microtype}
\usepackage{dcolumn}
\usepackage{graphicx}
\usepackage[small,bf,hang,raggedright,FIGTOPCAP]{subfigure}
\usepackage{varioref}
\usepackage{epstopdf}
\usepackage{color}
\usepackage{eucal}
\usepackage{float}
\usepackage{soul}
\usepackage{braket}
\usepackage{blkarray}
\usepackage{babel}
\usepackage{sansmath}
\usepackage{comment}
\usepackage{tikz,pgfplots,pgfplotstable}
\usepgfplotslibrary{external}
\usepgfplotslibrary{groupplots}
\usepgfplotslibrary{colorbrewer}
\usepgfplotslibrary{groupplots}
\pgfplotsset{cycle list/Set1}
\pgfplotsset{
    tick align=outside,
    tick pos=left,
    xmajorgrids,
    x grid style={white},
    ymajorgrids,
    y grid style={white},
    axis line style={white},
    axis background/.style={fill=white!89.803921568627459!black},
    legend style={draw=none, fill=none},
    legend cell align=left,
}
\pgfkeys{/pgf/number format/.cd, 1000 sep={}}

\tikzset{font=\small}

\begin{document}

\title{Phoebe: a High-Performance Framework for Solving Phonon and Electron Boltzmann Transport Equations}

\date{\today}

\author{Andrea Cepellotti}
\affiliation{Harvard John A. Paulson School of Engineering and Applied Sciences, Harvard University, Cambridge, MA 02138, USA}

\author{Jennifer Coulter}
\affiliation{Harvard John A. Paulson School of Engineering and Applied Sciences, Harvard University, Cambridge, MA 02138, USA}

\author{Anders Johansson}
\affiliation{Harvard John A. Paulson School of Engineering and Applied Sciences, Harvard University, Cambridge, MA 02138, USA}

\author{Natalya S. Fedorova}
\affiliation{Harvard John A. Paulson School of Engineering and Applied Sciences, Harvard University, Cambridge, MA 02138, USA}
\affiliation{Materials Research and Technology Department, Luxembourg Institute of Science and Technology,
5 Avenue des Hauts-Fourneaux, L-4362 Esch/Alzette, Luxembourg}

\author{Boris Kozinsky}
\email{bkoz@g.harvard.edu}
\affiliation{Harvard John A. Paulson School of Engineering and Applied Sciences, Harvard University, Cambridge, MA 02138, USA}
\affiliation{Robert Bosch LLC, Research and Technology Center, Cambridge MA}

\begin{abstract}
Understanding the electrical and thermal transport properties of materials is critical to the design of electronics, sensors and energy conversion devices.
Computational modeling can accurately predict materials properties but, in order to be reliable, require accurate descriptions of electron and phonon states and their interactions.
While first-principles methods are capable of describing the energy spectrum of each carrier, using them to compute transport properties is still a formidable task, both computationally demanding and memory intensive, requiring integration of fine microscopic scattering details for estimation of macroscopic transport properties.
To address this challenge, we present Phoebe -- a newly developed software package that includes the effects of electron-phonon, phonon-phonon, boundary, and isotope scattering in computations of electrical and thermal transport properties of materials with a variety of available methods and approximations.
This open source \verb_C++_ code combines MPI-OpenMP hybrid parallelization with GPU acceleration and distributed memory structures to manage computational cost, allowing Phoebe to effectively take advantage of contemporary computing infrastructures.
We demonstrate that Phoebe accurately and efficiently predicts a wide range of transport properties, opening avenues for accelerated computational analysis of complex crystals.
\end{abstract}

\maketitle

\section{Introduction}

As technology use and its energy footprint continue to grow, so does the importance of improving the properties of the materials which underlie such technology.
Materials transport properties, and in particular the capacity of a crystal to conduct heat and electricity, are critical to the function of virtually any electronic device.
These properties are determined by the spectrum and velocities of electrons and lattice vibrations in a crystal lattice, and the interactions of these carriers among themselves and with crystal defects.

The task of engineering better materials can be greatly accelerated through computational studies, as exemplified by successful computational discovery of new thermoelectric compounds \cite{doi:10.1146/annurev-matsci-100520-015716}.
In particular, the prediction of materials transport properties has reached high accuracy and robustness in recent years, thanks to the combination of first-principles density functional theory (DFT) simulations and numerical Boltzmann Transport Equation (BTE) solvers, the former capable of accurately parameterizing the microscopic characteristics of crystal quasiparticles and the latter providing macroscopic transport properties from microscopic data.
Both components are still the subject of study and are continuously evolving, as there are several challenges, both technical and conceptual, in both the calculation of the relevant microscopic scattering interactions, as well as in increasing the sophistication at which the BTE can be solved.
There are several active efforts on software implementations of these simulation techniques, and the challenge is to achieve the best possible accuracy for the least computational cost.
Various codes for solving the BTE are currently available, including EPW \cite{PONCE2016116}, BoltzWann \cite{PIZZI2014422}, PERTURBO \cite{ZHOU2021107970}, and BoltzTrap \cite{BoltzTraP2} for electronic properties, as well as phonopy/phono3py \cite{phonopy,phono3py}, ShengBTE \cite{LI20141747}, ALAMODE \cite{Alamode2014}, kALDo \cite{doi:10.1063/5.0020443} and GPU\_PBTE \cite{Zhang_2021} for the calculation of phonon transport.
The physics of heat and electricity conduction are intrinsically linked one to the other, and, as we will discuss in further detail, the electron and phonon BTE also share several common features. Despite the similarities, no single currently developed software package is capable of describing both kinds of transport processes.

Our vision behind the development of this new software was to take advantage of the parallels between physical descriptions of thermal and electrical conduction to provide a single framework to fully characterize materials transport properties.
In addition, we aimed not only to provide a tool for the prediction of materials transport properties; instead, we developed a package containing the abstraction and generic data structures needed to calculate conduction. In doing so, we created a flexible computational framework designed to easily accommodate extensions and improvements, as ongoing work in this field continues to reveal new and exciting developments.
Lastly, we capitalized on the rapid evolution of computational hardware, in particular with respect to the rise of GPU accelerators and the increasingly heterogeneous architecture of modern supercomputing facilities. 
As a result, we aspired to develop a new software which could run efficiently on any available computing environment.


This work presents the new open-source software package Phoebe -- a framework of PHOnon and Electron Boltzmann Equation solvers -- for the prediction of electron and phonon transport properties from first principles.
Phoebe implements the construction and solution of the Boltzmann transport equation for crystals using a scattering matrix formalism that takes into account the electron-phonon, phonon-phonon, boundary, and phonon-isotope scattering contributions to transport properties.
Phoebe implements several unique features, including Wigner transport formalism, the electron-phonon averaged approximation, and hydrodynamic transport properties. Additionally, the simultaneous   availability of both phonon and electron effects enables the calculation of properties such as the thermal conductivity of metals.

This code is designed to take advantage of contemporary supercomputing facilities to tackle the intensive computational workload of ab-initio transport calculations using multiple levels of parallelization.
The package is written in object-oriented \verb_C++_ and utilizes a hybrid MPI-OpenMP parallelization scheme, together with GPU acceleration, distributed memory data-structures, and parallel HDF5 input/output operations.
We discuss the capability of Phoebe to scale on supercomputing hardware and to reduce the computational cost of the calculation with respect to other codes, streamlining the calculation of transport properties.
With its flexible, object-oriented structure, Phoebe is designed for easy implementation of future features, such as additional quasiparticle interactions, and the possibility of interfacing with a wide range of ab-initio software for the calculation of interaction matrix elements. 

We showcase Phoebe's capabilities with the calculations of the transport properties of GaN, a III-V semiconductor used for high-power electronics.
After validating the results against experimental measurements, we argue that Phoebe is able to accurately and efficiently predict a wide range of material transport properties such as the electrical and thermal conductivity and thermoelectric performance in large and complex crystal structures. 

\section{Functionalities}

Phoebe provides the tools necessary for a complete first-principles study of electrical, thermal, and thermoelectric transport properties of crystalline materials in a single package. The code takes the output of ab-initio calculations of carrier spectra and coupling matrix elements and processes them in order to predict transport properties.
Phoebe provides the capability to compute a wide range of properties, such as:
\begin{itemize}
    \item Electron and phonon band-structure, density of states, and velocity operator;
    \item Linewidths and lifetimes of electrons, limited by electron-phonon interaction and boundary scattering; 
    \item Linewidths and lifetimes of phonons, limited by three-phonon, isotope and boundary scattering;
    \item Lattice thermal conductivity;
    \item Electrical conductivity, mobility, electronic thermal conductivity and Seebeck coefficient;
    \item Phonon and electron viscosity;
\end{itemize}
with more properties planned for future releases.
Moreover, we provide different approaches for modeling these properties at various levels of accuracy and/or computational cost. For example, the electron-phonon coupling can be evaluated either with the Wannier-interpolation technique, or the electron-phonon averaged approximation; transport properties can be evaluated using the BTE or the Wigner distribution function; and the BTE can be solved with various solvers, from relaxation time models to exact solutions.
Phoebe is made available freely on GitHub \cite{PhoebeRepo} under the MIT open-source license, and is written using object-oriented \verb_C++_, with the intention of allowing future extensions and implementation of new features with minimal effort.
We rely on a number of open-source libraries, including Eigen \cite{eigenweb} for small linear algebra problems and ScaLAPACK \cite{scalapack} for massively parallel linear algebra problems, as well as Spglib \cite{togo2018spglib} for analyzing symmetries, PugiXML \cite{pugiXML} for reading XML files, and HighFive \cite{HighFive} to handle I/O based using HDF5 \cite{hdf5}.

Additionally, Phoebe is designed to efficiently scale on contemporary HPC infrastructure.
Parallelization is achieved with a hybrid MPI and OpenMP architecture. The most computationally expensive parts of the code are further accelerated with Kokkos \cite{Kokkos}, which allows for the optional use of GPU accelerators.
Finally, we take advantage of several GitHub functionalities, such as GitHub pages to create a static website for user accessing the code and GitHub Actions to perform a continuous integration.
The documentation built with Doxygen \cite{doxygen} and Sphinx \cite{sphinx} is hosted online on Read The Docs \cite{readTheDocs}.


\section{Microscopic properties}
In this section, we report a brief summary of the main approaches used and implemented in Phoebe, that are needed to obtain a description of microscopic electron and phonon properties, referring to other works for detailed explanations.


\subsection{Phonon Properties}
\label{phononProperties}
In crystals, the motion of atoms is typically an oscillation around their equilibrium positions, and is thus described with the atomic displacements $u(s \alpha\boldsymbol{R})$, where $\boldsymbol{R}$ is the Bravais lattice vector labeling a unit cell, $s$ is an index over the unit cell basis of atoms, and $\alpha$ is the Cartesian direction of the atomic displacement.
To lowest orders in perturbation theory, the lattice potential energy near the minimum is described with a Taylor expansion \cite{ziman}, where the first two important terms are the second derivative $V^{(2)}$ of the crystal energy with respect to atomic displacements
\begin{equation}
V^{(2)}(s\alpha\boldsymbol{R}, s'\beta\boldsymbol{R}') 
= 
\frac{\partial^2 E} { \partial u(s \alpha\boldsymbol{R}) \partial u(s' \beta\boldsymbol{R}') } \;,
\end{equation}
and the third derivative $V^{(3)}$
\begin{equation}
V^{(3)}(s\alpha\boldsymbol{R},s'\beta\boldsymbol{R}',s''\gamma\boldsymbol{R}'') 
= 
\frac{\partial^3 E} { \partial u(s \alpha\boldsymbol{R}) \partial u(s' \beta\boldsymbol{R}') \partial u(s'' \gamma\boldsymbol{R}'')  } \;,
\end{equation}
while higher order derivatives are here neglected.
These quantities can be computed using existing ab-initio codes.
In particular, Phoebe interfaces with phonopy/phono3py \cite{phonopy} and Quantum ESPRESSO (QE) \cite{QE-2017} for the calculation of the second derivative and with phono3py \cite{phono3py} and ShengBTE \cite{ShengBTE_2014} for the calculation of the third derivative of the energy.

These two derivatives allow the characterization of phonon properties and of their scattering mechanisms by means of a Fourier transform.
For example, the Fourier transform of the second derivative provides the dynamical matrix $D$ defined as \cite{RevModPhysBaroni}
\begin{equation}
\label{phononInterpolation}
      D(s\alpha | s'\alpha')(\boldsymbol{q}) = \sum_{\boldsymbol{R}} V^{(2)}(s\alpha \boldsymbol{0}| s'\alpha' \boldsymbol{R}) e^{i \boldsymbol{q} \cdot \boldsymbol{R}} \;,
\end{equation}
where $\boldsymbol{q}$ is an arbitrary wavevector, and one Bravais lattice vector component of the second derivative $V^{(2)}$ has been set to a reference unit cell $\boldsymbol{0}$ without loss of generality thanks to the translational periodicity of the crystal.
The phonon harmonic properties are then found by diagonalizing the dynamical matrix
\begin{equation}
\label{phononEnergyDefinition}
   D(s\alpha | s'\alpha')(\boldsymbol{q}) z_{s'\alpha'\nu}(\boldsymbol{q}) = (\hbar\omega_{\nu}(\boldsymbol{q}))^2  z_{s\alpha \nu}(\boldsymbol{q}) \;,
\end{equation}
where $\nu$ is an eigenvalue index (i.e. the phonon branch), $\hbar \omega_{\boldsymbol{q}\nu}$ the phonon energy and $z_{\boldsymbol{q}\nu}^{s\alpha}$ the phonon eigenvector.
Polar corrections to the dynamical matrix are taken into account in Phoebe as explained in Ref. \cite{RevModPhysBaroni}.

The Fourier transform of the third derivative $V^{(3)}$ instead describes the coupling strength of a 3-phonon scattering event, and is computed as \cite{fugalloVariational,PaulattoAnharmonic}
\begin{equation}
\label{3PhCouplingInterpolation}
    V^{(3)}(\boldsymbol{q} \nu,\boldsymbol{q}' \nu',\boldsymbol{q}'' \nu'') 
    = 
    \frac{1}{N_0^3}
    \left(\frac{2}{\hbar}\right)^{3/2}
    \sum_{\boldsymbol{R}\boldsymbol{R}'}
    \sum_{\alpha\beta\gamma}
    \sum_{ss's''}
    \sqrt{\frac{
    m_{s}m_{s'}m_{s''}}{
    \omega_{\boldsymbol{q}\nu}   \omega_{\boldsymbol{q}'\nu'}  \omega_{\boldsymbol{q}''\nu''}
    }}
    z^{s \alpha^*}_{\boldsymbol{q} \nu}
    z^{s' \beta^*}_{\boldsymbol{q}' \nu'}
    z^{s'' \gamma^*}_{\boldsymbol{q}'' \nu''}
    e^{-i (\boldsymbol{q} \cdot \boldsymbol{R} + \boldsymbol{q}' \cdot \boldsymbol{R}')}
    V^{(3)}(s\alpha\boldsymbol{R},s'\beta \boldsymbol{R}',s''\gamma \boldsymbol{0})  \;,
\end{equation}
where again one Bravais lattice vector component can be set to a reference unit cell $\boldsymbol{0}$.

The equations \ref{phononInterpolation} and \ref{3PhCouplingInterpolation} allow us to compute phonon properties (such as phonon energies or scattering coupling strength) at arbitrary wavevectors, starting from the derivatives computed in real space that can be obtained from ab-initio codes.
This interpolation procedure is a major computational bottleneck due to the large number of tensor indices and the fact that phonon properties need to be computed for all wavevectors used to sample the Brillouin zone (BZ).
As a result, it is critical to efficiently implement this interpolation procedure, in particular for the 3-phonon coupling strength, as will be discussed in Section \ref{Section:Kokkos}.


\subsection{Electron Properties}
\label{Section:ElectronProperties}
In order to compute the electronic transport properties, it is necessary to characterize electrons and the electron-phonon interaction.
Phoebe relies on an external ab-initio code to compute the electronic energies $\epsilon_{\boldsymbol{k}b}$ and wavefunctions $\ket{\psi_{\boldsymbol{k}b}}$, where $\boldsymbol{k}$, the electron wavevector, and $b$, the band index, are Bloch quantum numbers (spin is omitted for simplicity), which correspond to eigenvalues and eigenstates of the electronic Hamiltonian
\begin{equation}
    H^B(\boldsymbol{k}) \ket{\psi_{\boldsymbol{k}b}} = \epsilon_{\boldsymbol{k}b} \ket{\psi_{\boldsymbol{k}b}} \;.
\end{equation}
Once the electronic wavefunction is known, it is possible to compute other properties such as the interaction of electrons with the lattice.
The electron-phonon coupling $g$ is defined as \cite{RevModPhys.89.015003}
\begin{equation}
\label{elphCouplingDefinition}
    g_{b'b\nu}(\boldsymbol{k},\boldsymbol{q})
    =
    \left(\frac{1}{2m\omega_{\boldsymbol{q}\nu}}\right)^{1/2}
    \bra{\psi_{\boldsymbol{k}+\boldsymbol{q},b'}}
    \partial_{\boldsymbol{q}\nu} V
    \ket{ \psi_{\boldsymbol{k}b}}
    =
    \left(\frac{1}{2m\omega_{\boldsymbol{q}\nu}}\right)^{1/2}
    g^0_{b'b\nu}(\boldsymbol{k},\boldsymbol{q}) \;,
\end{equation}
and provides the coupling strength of the interaction between two electrons and a phonon, and $\partial_{\boldsymbol{q}\nu} V$ is the potential originating in response to the perturbation of the lattice with the ionic displacements of a phonon mode $\boldsymbol{q}\nu$. 
We define $g^0$ here as a quantity smoother to interpolate than $g$, as will be discussed below. 

In this initial version, Phoebe interfaces with QE \cite{QE-2017} in order to obtain values of the electronic energies and the electron-phonon coupling. In principle, ab-initio codes can directly compute electronic properties at any wavevector $\boldsymbol{k}$.
The calculation of transport properties using momentum-space integrals typically requires very fine sampling of $\boldsymbol{k}$ throughout the Brillouin zone, making a direct first-principles calculation of every state unfeasible in most cases.
To circumvent this issue, transport calculations rely on an interpolation scheme, in which QE computes electronic properties on a coarse grid of wavevectors, $\{\boldsymbol{k}_{co}\}$, and Phoebe interpolates these properties to an arbitrary fine mesh of wavevectors $\{\boldsymbol{k}_{fi}\}$. 

We implemented the Wannier interpolation technique, interfacing Phoebe with Wannier90 \cite{Wannier90} for the construction of maximally localized Wannier functions (MLWF) \cite{RevModPhysMarzari, WannierInterpolation}.
The Wannier function $\ket{\boldsymbol{R}m}$ computed by Wannier90 is defined as
\begin{equation}
    \ket{\boldsymbol{R}m}
    =
    \sum_{b\boldsymbol{k}_{co}}
    e^{-i\boldsymbol{k}_{co}\cdot \boldsymbol{R}}
    U_{\boldsymbol{k}_{co},mb}
    \ket{\psi_{\boldsymbol{k}_{co}b}} \;,
\end{equation}
where $m$ is a Wannier function index, and $U$ is a rotation matrix.
To briefly explain the MLWF procedure, we first recall that the wavefunction $\ket{\psi_{\boldsymbol{k}b}}$ is not unique, as it has a gauge degree of freedom: the wavefunction can be multiplied by an arbitrary factor $e^{i\phi(\boldsymbol{k},b)}$.
In practice, the phase $\phi(\boldsymbol{k},b)$ is a random result of the numerical diagonalization of a Hamiltonian matrix, and depends unpredictably on the linear algebra algorithm implementation.  
As a result of this phase arbitrariness, the Wannier function are not uniquely defined.
The MLWF procedure aims at reducing this arbitrariness, by finding for a rotation matrix $U$ such that the spread of the Wannier function $\ket{\boldsymbol{R}m}$ is minimized.

The interpolation of electronic energies proceeds as follows \cite{WannierInterpolation}: we first compute the Hamiltonian matrix in the basis of Wannier functions $\langle \boldsymbol{0} n | H | \boldsymbol{R} m \rangle$ as
\begin{equation}
\langle \boldsymbol{0} n | H | \boldsymbol{R} m \rangle
= 
\frac{1}{N_k}
\sum_{\boldsymbol{k}_{co},bb'} e^{-i\boldsymbol{k}_{co} \cdot \boldsymbol{R}}
U_{\boldsymbol{k}_{co},nb} [ \delta_{bb'} \epsilon_{\boldsymbol{k}_{co}b} ] U^{\dagger}_{\boldsymbol{k}_{co},mb'} \;,
\end{equation}
where $N_k$ is the total number of wavevectors in the coarse grid $\{\boldsymbol{k}_{co}\}$.
From the Hamiltonian in the Wannier representation, we interpolate electronic energies at any arbitrary wavevector $\boldsymbol{k}$ by computing a Fourier back-transform
\begin{equation}
       H_{\boldsymbol{k},nm}^W = \sum_{\boldsymbol{R}} e^{i \boldsymbol{k} \cdot \boldsymbol{R}}  \langle \boldsymbol{0} n | H | \boldsymbol{R} m \rangle \;,
\end{equation}
followed by a diagonalization of the resulting matrix
\begin{equation}
   H_{\boldsymbol{k},bb'}^B = [U_{\boldsymbol{k}}^\dagger H_{\boldsymbol{k}}^W U_{\boldsymbol{k}}]_{bb'} = \delta_{bb'} \epsilon_{\boldsymbol{k}b} \;,
\end{equation}
which provides an interpolation of the electronic energies $\epsilon_{\boldsymbol{k}b}$ and rotation matrix $U_{\boldsymbol{k}}$ at any desired wavevector.

The electron-phonon coupling can be interpolated in a conceptually similar way to the electronic energies.
Following the approach of Ref. \cite{PhysRevB.76.165108}, we first note that the matrix $g^0$ is smoother than $g$ (as defined in Eq. \ref{elphCouplingDefinition}) in the reciprocal space, since it removes a dependence on the phonon dispersion relation, and $g^0$ is thus more suitable for an interpolation algorithm.
The matrix $g^0(\boldsymbol{k}_{co},\boldsymbol{q}_{co})$ (with band indices omitted for simplicity) is computed by QE on a coarse grid of wavevectors and is then transformed by Phoebe to the Wannier representation as
\begin{equation}
\label{blochToWannierG}
   g^0(\boldsymbol{R}_e,\boldsymbol{R}_p)
   =
   \frac{1}{N_q N_k}
   \sum_{\boldsymbol{k}_{co}\boldsymbol{q}_{co}} e^{-i\boldsymbol{k}_{co}\cdot\boldsymbol{R}_e-i\boldsymbol{q}_{co}\cdot\boldsymbol{R}_p} U_{\boldsymbol{k}_{co}+\boldsymbol{q}}^\dagger g(\boldsymbol{k},\boldsymbol{q}) U_{\boldsymbol{k}_{co}} z_{\boldsymbol{q}_{co}}^{-1} \;,
\end{equation}
where $N_q$ is the number of phonon wavevectors in the coarse sampling, $z$ is the phonon eigenvector introduced in Eq. \ref{phononEnergyDefinition}, and $\boldsymbol{R}_e$ ($\boldsymbol{R}_p$) is the set of Bravais lattice vectors associated to the electron (phonon) Fourier transform of the set $\{\boldsymbol{k}_{co}\}$ ($\{\boldsymbol{q}_{co}\}$).
Once this transformation is performed, the electron-phonon matrix is finally interpolated to arbitrary wavevectors $\boldsymbol{k}$ and $\boldsymbol{q}$ as
\begin{equation}
\label{elPhCouplingInterpolation}
   g^0(\boldsymbol{k},\boldsymbol{q})
   =
   \frac{1}{N_e}
   \sum_{\boldsymbol{R}_e \boldsymbol{R}_p} e^{i\boldsymbol{k}\cdot\boldsymbol{R}_e+i\boldsymbol{q}\cdot\boldsymbol{R}_p} U_{\boldsymbol{k}+\boldsymbol{q}} g^0(\boldsymbol{R}_e,\boldsymbol{R}_p) U_{\boldsymbol{k}}^\dagger z_{\boldsymbol{q}} \;,
\end{equation}
where the $U$ and $z$ matrices at the interpolating wavevectors can be found by constructing the dynamical matrix at the desired $\boldsymbol{q}$ wavevector and by interpolating the electronic energy at wavevector $\boldsymbol{k}$.
As a result of this interpolation procedure, one can accurately sample the the electron-phonon coupling over the BZ and compute transport properties.

Similar to the interpolation of the three-phonon coupling, the interpolation of the electron-phonon coupling is a computationally expensive effort due to the large number of indices and the sheer number of interpolating wavevectors often needed to converge transport integrals.
The implementation of the coupling interpolation for both electron-phonon and three-phonon coupling will be discussed together in detail in Section \ref{Section:Kokkos}, as Eq. \ref{3PhCouplingInterpolation} and Eq. \ref{elPhCouplingInterpolation} are similar from a mathematical point of view.

Finally, polar corrections to the electron-phonon coupling are implemented in Phoebe following the procedure outlined in Ref. \cite{PhysRevLett.115.176401}. The coupling is decomposed into short- and long-range contributions,
\begin{equation}
    g_{bb'\nu}(\boldsymbol{k},\boldsymbol{q}) = 
    g_{bb'\nu}^S(\boldsymbol{k},\boldsymbol{q}) +
    g_{bb'\nu}^L(\boldsymbol{k},\boldsymbol{q}), \label{eq:gdecouple}
\end{equation}
where the long-range contribution is given by
\begin{equation}
    g_{b'b\nu}^L(\boldsymbol{k},\boldsymbol{q})
    = i\frac{4\pi}{V}\frac{e^2}{4\pi\epsilon_0}
    \sum_s \left(\frac{\hbar}{2N_q m_s\omega_{\boldsymbol{q}\nu}}\right)^{1/2}
    \sum_{\boldsymbol{G}\neq\boldsymbol{q}}
    \frac{\left(\boldsymbol{q}+\boldsymbol{G}\right)\cdot \boldsymbol{Z}_s^*\cdot \boldsymbol{z}_{\boldsymbol{q}\nu}^s}
    {\left(\boldsymbol{q}+\boldsymbol{G}\right)\cdot\boldsymbol{\epsilon}^\infty\cdot \left(\boldsymbol{q}+\boldsymbol{G}\right)}
    \left\langle \psi_{\boldsymbol{k}+\boldsymbol{q},b'}\left| e^{i\left(\boldsymbol{q}+\boldsymbol{G}\right)\cdot\boldsymbol{r}} \right|\psi_{\boldsymbol{k}+\boldsymbol{q},b}\right\rangle,\label{eq:polarlong}
\end{equation} 
and the short-range contribution is calculated on the coarse grid by subtracting the long-range term from the total electron-phonon coupling calculated by the ab-initio code.
Here, \(\boldsymbol{Z}_s^*\) is the Born effective charge tensor of atom \(s\), and \(\epsilon_0\) and \(\boldsymbol{\epsilon}^\infty\) are the low- and high-frequency permittivities.


\subsection{Velocity}
For charge carriers, we compute the matrix elements of the velocity operator $v$ as the derivative with respect to the wavevector of the non-interacting Hamiltonian \cite{BLOUNT1962305}
\begin{equation}
    v_{\alpha \boldsymbol{k},bb'}
    =
    \sum_{mn}
    U^{\dagger}_{\boldsymbol{k},mb} \frac{\partial H^W_{\boldsymbol{k},mn} }{\partial k_\alpha} U_{\boldsymbol{k},nb'} 
    \approx 
    \frac{1}{2 \delta_\alpha} \sum_{mn}
    U^{\dagger}_{\boldsymbol{k},mb} \Big( H^W_{\boldsymbol{k+\delta_\alpha},mn} -
    H^W_{\boldsymbol{k-\delta_\alpha},mn} \Big) U_{\boldsymbol{k},nb'} \;,
\end{equation}
where $H^W$ is the Hamiltonian matrix in the Bloch representation computed at $\boldsymbol{k}$ with the Wannier interpolation technique described above, and $\delta_\alpha$ is a small displacement vector along Cartesian direction $\alpha$.
If the eigenvalue $\epsilon_{\boldsymbol{k}b}$ is non-degenerate, the diagonal matrix element $v_{\alpha\boldsymbol{k},bb}$ represents the electronic group velocity.
However if $\epsilon_{\boldsymbol{k}b}$ is a degenerate eigenvalue, we cannot readily identify the velocity matrix element $v_{\alpha \boldsymbol{k},bb}$ with the group velocity: in this case, we select the  degenerate subspace of the matrix $v_{\alpha\boldsymbol{k}}$ and diagonalize it.
After this subspace diagonalization, the new diagonal matrix elements represent the group velocity of the degenerate state.

The definition of the phonon velocity is similar to that of electrons, allowing us to use the same code for both velocity calculations.
Simply, the matrix $H$ is replaced with the square root of the dynamical matrix $\sqrt{D_{\boldsymbol{q}}}$ and the rotation matrix $U$ is replaced with the phonon eigenvector $z$ \cite{fugalloVariational}.


\section{Transport Methods}
Moving on to the transport properties, we show in this section how the microscopic properties described above can be used to evaluate macroscopic transport coefficients.

\subsection{Phonon Boltzmann Transport Equation}
\label{section:phBTE}
Given the microscopic quantities discussed in the previous sections, we discuss now how to obtain the lattice thermal conductivity.
We describe lattice thermal transport in the framework of the phonon Boltzmann transport equation (BTE) \cite{peierls, ziman}.
This semiclassical model describes the motion of a phonon (wavepacket) through the crystal.
A phonon state $(\boldsymbol{q},\nu)$ at thermal equilibrium is populated with $\bar{n}_{\boldsymbol{q}\nu}$ phonon quanta, where $\bar{n}_{\boldsymbol{q}\nu} = 1/(e^{\hbar \omega_{\boldsymbol{q}\nu}/k_BT}-1)$ is the Bose--Einstein distribution.
The application of a thermal gradient $\nabla T$ to the crystal forces the system out-of-equilibrium, so that each phonon state is then populated with an out-of-equilibrium population $n_{\boldsymbol{q}\nu}$.
The phonon BTE provides a way to find such out-of-equilibrium population and consists of a balance between two terms:
\begin{equation}
\label{phBte}
    \frac{\partial \bar{n}_{\boldsymbol{q}\nu}} {\partial T}
    v_{\alpha \boldsymbol{q}\nu} \nabla_\alpha T  =
    - \frac{1}{V N_q} \sum_{\boldsymbol{q}'\nu'} \Omega_{\boldsymbol{q}\nu, \boldsymbol{q}'\nu'} \delta n_{\alpha \boldsymbol{q}'\nu'} \;.
\end{equation}
The left-hand side is the diffusion operator, and describes the phonon drift in presence of a thermal gradient $\nabla T$, while the right-hand side describes how scattering changes the population number of a phonon state.
Here, $V$ is the crystal unit cell volume, $\Omega_{\boldsymbol{q}\nu,\boldsymbol{q}'\nu'}$ is the scattering matrix and $\delta n_{\boldsymbol{q}\nu} = n_{\boldsymbol{q}\nu} - \bar{n}_{\boldsymbol{q}\nu}$ is the deviation from equilibrium population.
For computational convenience, we also work with a rescaled scattering matrix $A$ defined as $A_{\boldsymbol{q}\nu,\boldsymbol{q}'\nu'}
=
\Omega_{\boldsymbol{q}\nu,\boldsymbol{q}'\nu'} \bar{n}_{\boldsymbol{q}'\nu'}(\bar{n}_{\boldsymbol{q}'\nu'}+1)$.
The phonon scattering matrix is built taking into account the three-phonon interaction, phonon-isotope, phonon-boundary and phonon-electron scattering \cite{fugalloVariational} as
\begin{align}
\label{phScatteringMatrix}
   A_{\boldsymbol{q}\nu,\boldsymbol{q}'\nu'} = & \left[{\sum_{\boldsymbol{q}''\boldsymbol{q}'''\nu''\nu'''}} \left(P^{\boldsymbol{q}''\nu''}_{\boldsymbol{q}\nu,\boldsymbol{q}'''\nu'''} + \frac{ P_{\boldsymbol{q}'''\nu''',\boldsymbol{q}''\nu''}^{\nu}}{2} \right) 
   + \sum_{\boldsymbol{q}''\nu''} P^{\mathrm{isot}}_{\boldsymbol{q}\nu,\boldsymbol{q}''\nu''} 
   + P^{\mathrm{be}}_{\boldsymbol{q}\nu}
   + P^{\mathrm{el-ph}}_{\boldsymbol{q}\nu}
   \right]
   \delta_{\nu,\nu'}
   \delta_{\boldsymbol{q},\boldsymbol{q}'} \\
   &- {\sum_{\boldsymbol{q}''\nu''}} \left(  P^{\boldsymbol{q}'\nu'}_{\boldsymbol{q}\nu,\boldsymbol{q}''\nu''} -P^{\boldsymbol{q}''\nu''}_{\boldsymbol{q}\nu,\boldsymbol{q}'\nu'}+ P_{\boldsymbol{q}'\nu',\boldsymbol{q}''\nu''}^{\boldsymbol{q}\nu}  \right) + P^{\mathrm{isot}}_{\boldsymbol{q}\nu,\boldsymbol{q}'\nu'} \nonumber \;, 
\end{align}
where we introduced the three-phonon scattering rates for coalescence events as
\begin{equation}
   P^{\boldsymbol{q}'' \nu''}_{\boldsymbol{q} \nu,\boldsymbol{q}' \nu'} 
   = 
   \frac{2 \pi}{N_q \hbar^2} \sum_{\boldsymbol{G}}
   \left|V^{(3)}(\boldsymbol{q} j,\boldsymbol{q}' \nu',-\boldsymbol{q}'' \nu'')\right|^2
   \bar{n}_{\boldsymbol{q} j}\bar{n}_{\boldsymbol{q}' \nu'}(\bar{n}_{\boldsymbol{q}'' \nu''}+1) \delta_{\boldsymbol{q}+\boldsymbol{q}' -\boldsymbol{q}'', \boldsymbol{G}}
   \delta(\hbar \omega_{\boldsymbol{q} \nu} +\hbar \omega_{\boldsymbol{q}' \nu'}-\hbar \omega_{\boldsymbol{q}'' \nu''}) \;,
\end{equation}
while the scattering rate for phonon decay is
\begin{equation}
   P^{\boldsymbol{q}' \nu',\boldsymbol{q}'' \nu''}_{\boldsymbol{q} \nu} = \frac{2 \pi}{N_q \hbar^2 } \sum_{\boldsymbol{G}}
   \left|V^{(3)}(\boldsymbol{q} \nu,-\boldsymbol{q}' \nu',-\boldsymbol{q}'' \nu'')\right|^2
   \bar{n}_{\boldsymbol{q} \nu} (\bar{n}_{\boldsymbol{q}' \nu'}+1) (\bar{n}_{\boldsymbol{q}'' \nu''}+1) \delta_{\boldsymbol{q}-\boldsymbol{q}' -\boldsymbol{q}'', \boldsymbol{G}}
   \delta(\hbar \omega_{\boldsymbol{q} \nu}-\hbar \omega_{\boldsymbol{q}' \nu'}-\hbar \omega_{\boldsymbol{q}'' \nu''} ) \;,
\end{equation}
where $\boldsymbol{G}$ is a reciprocal lattice vector, $N_q$ is the number of phonon wavevectors for the BZ integration, and $V^{(3)}(\boldsymbol{q} \nu,\boldsymbol{q}' \nu',\boldsymbol{q}'' \nu'')$ is the three-phonon coupling strength defined in Eq. (\ref{3PhCouplingInterpolation}).

The phonon-electron interaction contributes to the scattering matrix with a term on the diagonal part \cite{liao2015significant}
\begin{equation}
P^{\mathrm{el-ph}}_{\boldsymbol{q} \nu}
=
- \frac{2\pi}{N_{\boldsymbol{k}}\hbar} 
\bar{n}_{\boldsymbol{q} \nu}(\bar{n}_{\boldsymbol{q} \nu}+1)
\sum_{\boldsymbol{k}bb'}  \left|g_{bb'\nu}(\boldsymbol{k},\boldsymbol{q})\right|^{2} \left(f_{\boldsymbol{k}b}-f_{\boldsymbol{k}+\boldsymbol{q}b'}\right) \delta\left(\varepsilon_{\boldsymbol{k}b}-\varepsilon_{ \boldsymbol{k}+\boldsymbol{q}b'}-\omega_{\boldsymbol{q} \nu}\right) \;,
\end{equation}
with $f_{\boldsymbol{k}b}-f_{\boldsymbol{k}+\boldsymbol{q}b'} \approx \frac{\partial f_{\boldsymbol{k}b}}{\partial \varepsilon_{ \boldsymbol{k}b}} \hbar \omega_{\boldsymbol{q} \nu}$ for numerical stability.
Additionally, the scattering rate for phonon-isotope scattering is \cite{marzari:isotopes}
\begin{equation}
   P_{\boldsymbol{q} \nu,\boldsymbol{q}' \nu'}^{\mathrm{isot}} = \frac{\pi}{2 N_q} \omega_{\boldsymbol{q} \nu}\omega_{\boldsymbol{q}' \nu'}
   \left[ \bar{n}_{\boldsymbol{q} \nu} \bar{n}_{\boldsymbol{q}' \nu'} + \frac{\bar{n}_{\boldsymbol{q} \nu} + \bar{n}_{\boldsymbol{q}' \nu'}} {2} \right ]
   \sum_{s} g^{isot}  \left|  \sum_{\alpha} z^{s \alpha^*}_{\boldsymbol{q}\nu} \cdot z^{s \alpha}_{\boldsymbol{q}' \nu'} \right|^2 \delta (\omega_{\boldsymbol{q} \nu}- \omega_{\boldsymbol{q}' \nu'}) \;,
\end{equation}
where the coupling strength $g^{isot}_s$ depends on the variance of masses as 
\begin{equation}
   g^{isot}_s = \frac{1}{\langle m_s \rangle^2}
   \sum_{i} f_{is} ( m_{is} - \langle m_s \rangle )^2 \;,
\end{equation}
where $m_{is}$ is the atomic mass of atom of species $s$ and isotope $i$, $f_{is}$ is the abundance of the isotope (which by default is set to natural abundances \cite{isotopesAbundance}) and $\langle m_s \rangle$ is the average atomic mass.
Finally, the boundary scattering rate is described with the term
\begin{equation}
   P_{\boldsymbol{q} \nu}^{\mathrm{be}} = \frac{v_{\boldsymbol{q} \nu}}{L} \bar{n}_{\boldsymbol{q} \nu} (\bar{n}_{\boldsymbol{q} \nu}+1),
\end{equation}
where $L$ is the user-specified macroscopic crystal size.

Using the first-principles phonon properties discussed in Section \ref{phononProperties} it is possible to construct the phonon BTE and in particular the scattering matrix.
In Section \ref{Section:BteSolvers}, we will discuss how to solve the BTE to obtain the out-of-equilibrium population $n_{\boldsymbol{q}\nu}$.
Assuming that $n_{\boldsymbol{q}\nu}$ is found, the thermal conductivity tensor $\kappa_{\alpha\beta}$ is computed as
\begin{equation}
    \kappa_{\alpha\beta} = \frac{1}{N_q V} \sum_{\boldsymbol{q}\nu} \hbar \omega_{\boldsymbol{q}} v_{\alpha\boldsymbol{q}\nu}
    \bigg( \frac{ \delta n_{\boldsymbol{q}\nu} }{ \nabla T} \bigg)_{\beta} \;,
\end{equation}
where $v_{\alpha\boldsymbol{q}\nu}$ is the phonon group velocity.


\subsection{Electron Boltzmann Transport Equation}
\label{Section:ElectronBTE}
The semiclassical description of electron transport shares several similarities with the phonon transport theory.
In the phonon case, we solved for out-of-equilibrium phonon populations. Now, for the case of electron transport, we deal with the problem of finding the out-of-equilibrium occupation number of a charge carrier.
At equilibrium, an electronic state $(\boldsymbol{k},b)$ is occupied according to the Fermi--Dirac distribution $f_{\boldsymbol{k}b} = 1/(e^{(\epsilon_{\boldsymbol{k}b} - \mu)/k_BT} + 1)$, where $\mu$ is the chemical potential and $\epsilon_{\boldsymbol{k}b}$ is the electronic energy.
When we apply a thermal gradient to a material, we can expect to induce the diffusion of carriers just as in the phonon case, resulting in an out-of-equilibrium population $f_{\boldsymbol{k}b}$.
However, each carrier also possesses a charge $e$, and therefore can diffuse also when an external electric field $\boldsymbol{E}$ is applied to the material.
As we are only concerned with the linear response to the perturbation induced by the thermal gradient or the electric field, the electronic BTE consists of two separate problems for the two different perturbations $\boldsymbol{E}$ and $\nabla T$.
The BTE describing the response to an electric field in direction $\alpha$ is
\begin{equation}
\label{elBte1}
    \frac{\partial  \bar{f}_{\boldsymbol{k}b}} {\partial \epsilon}
    v_{\alpha \boldsymbol{k}b} e E_{\alpha} 
    =
    - \frac{1}{N_k V} \sum_{\boldsymbol{k}'b'} \Omega_{\boldsymbol{k}b, \boldsymbol{k}'b'} \delta f_{\alpha \boldsymbol{k}'b'} \;,
\end{equation}
where $\delta f_{\boldsymbol{k}b} = f_{\boldsymbol{k}b} - \bar{f}_{\boldsymbol{k}b}$ is the deviation from the equilibrium population, $v_{\alpha\boldsymbol{k}b}$ is the electronic group velocity in direction $\alpha$, and $\Omega$ is the electronic scattering matrix.
The BTE describing the response to a thermal gradient perturbation closely resembles the phonon case and is
\begin{equation}
\label{elBte2}
    \frac{\partial 
    \bar{f}_{\boldsymbol{k}b}}{\partial T}
    v_{\alpha \boldsymbol{k}b}  \nabla_{\alpha} T =
    - \frac{1}{N_k V} \sum_{\boldsymbol{k}'b'} \Omega_{\boldsymbol{k}b, \boldsymbol{k}'b'} \delta f_{\alpha \boldsymbol{k}'b'} \;.
\end{equation}

When taking the electron-phonon scattering into account, the electron scattering matrix $\Omega$ can be computed as \cite{bonini2016}
\begin{align}
\label{elScatteringMatrix}
  \Omega_{\boldsymbol{k}b, \boldsymbol{k}'b'}
  =
  \frac{1}{\tau_{\boldsymbol{k}b}} \delta_{\boldsymbol{k}b,\boldsymbol{k}'b'}
  +
  (1-\delta_{\boldsymbol{k}b,\boldsymbol{k}'b'})
  \frac{2\pi}{N_k V}
  \sum_{\boldsymbol{q}\nu}
  |g_{bb'\nu}(\boldsymbol{k},\boldsymbol{k}')|^2
  \Big[ & (1-\bar{f}_{\boldsymbol{k}'b'} + \bar{n}_{\boldsymbol{q}\nu}) \delta(\epsilon_{\boldsymbol{k}b}-\hbar\omega_{\boldsymbol{q}\nu}-\epsilon_{\boldsymbol{k}'b'}) \\
   &+
  (\bar{f}_{\boldsymbol{k}'b'} + \bar{n}_{\boldsymbol{q}\nu}) \delta(\epsilon_{\boldsymbol{k}b}+\hbar\omega_{\boldsymbol{q}\nu}-\epsilon_{\boldsymbol{k}'b'})
  \Big]
  \delta(\boldsymbol{k}-\boldsymbol{k}'+\boldsymbol{q}) \;, \nonumber
\end{align}
where the lifetime $\tau_{\boldsymbol{k}b}$ is
\begin{equation}
    \frac{1}{\tau_{\boldsymbol{k}b}}
    = 
    \frac{2\pi}{N_k V}
  \sum_{\boldsymbol{k}'b' \boldsymbol{q}\nu}
  |g_{bb'\nu}(\boldsymbol{k},\boldsymbol{k}')|^2
  \Big[ (1-f_{\boldsymbol{k}'b'} + \bar{n}_{\boldsymbol{q}\nu})
  \delta(\epsilon_{\boldsymbol{k}b}-\hbar\omega_{\boldsymbol{q}\nu}-\epsilon_{\boldsymbol{k}'b'})
  +
  (f_{\boldsymbol{k}'b'} + \bar{n}_{\boldsymbol{q}\nu}) \delta(\epsilon_{\boldsymbol{k}b}+\hbar\omega_{\boldsymbol{q}\nu}-\epsilon_{\boldsymbol{k}'b'})
  \Big]
  \delta(\boldsymbol{k}-\boldsymbol{k}'+\boldsymbol{q})
\end{equation}
where $g_{bb'\nu}(\boldsymbol{k},\boldsymbol{k}')$ is the coupling strength of the electron-phonon interaction.
Electron scattering against surfaces or boundaries can be taken into account by adding to $A$ a term of the form $\frac{v_{\alpha \boldsymbol{k}b}}{L} \delta_{\boldsymbol{k},\boldsymbol{k}'} \delta_{b,b'}$, where $L$ is a characteristic length. 

We note that the scattering matrix $\Omega$ can be brought to a symmetric form with the scaling
\begin{equation}
    \tilde{\Omega}_{\boldsymbol{k}b, \boldsymbol{k}'b'} 
    =
    \Omega_{\boldsymbol{k}b, \boldsymbol{k}'b'}
    \sqrt{ \frac{\bar{f}_{\boldsymbol{k}'b'}(1-\bar{f}_{\boldsymbol{k}'b'})}{\bar{f}_{\boldsymbol{k}b}(1-\bar{f}_{\boldsymbol{k}b})} } 
\end{equation}
it can readily be checked that the matrix $\tilde{\Omega}$ is symmetric (i.e. $\tilde{\Omega}_{\boldsymbol{k}b, \boldsymbol{k}'b'} = \tilde{\Omega}_{\boldsymbol{k}'b', \boldsymbol{k}b} $), since the matrix elements are
\begin{align}
    \tilde{\Omega}_{\boldsymbol{k}b, \boldsymbol{k}'b'} = \frac{1}{\tau_{\boldsymbol{k}b}} \delta_{\boldsymbol{k}b\boldsymbol{k}'b'} 
    +
    (1-\delta_{\boldsymbol{k}b,\boldsymbol{k}'b'})
    \frac{2\pi}{N_k V}
  \sum_{\boldsymbol{k}'b' \boldsymbol{q}\nu}
  |g_{bb'\nu}(\boldsymbol{k},\boldsymbol{k}')|^2
  & \Big[ 
  \delta(\epsilon_{\boldsymbol{k}b}-\hbar\omega_{\boldsymbol{q}\nu}-\epsilon_{\boldsymbol{k}'b'}) 
  +
  \delta(\epsilon_{\boldsymbol{k}b}+\hbar\omega_{\boldsymbol{q}\nu}-\epsilon_{\boldsymbol{k}'b'})
  \Big] \\
  & \times \delta(\boldsymbol{k}-\boldsymbol{k}'+\boldsymbol{q})
  \frac{1}{2\sinh(\frac{\hbar\omega_{\boldsymbol{q}\nu}}{2k_BT})};. \nonumber
\end{align}
This symmetric expression for the scattering matrix is more suitable for implementing linear algebra solvers to the BTE, and has been developed in analogy to the rescaling of the phonon scattering matrix described in Ref. \cite{relaxons}.

The solution to the electronic BTE is discussed in Section \ref{Section:BteSolvers}. Solving the BTE provides a way of calculating the linear response of the electronic out-of-equilibrium population in response to an electric field $\delta f^{E\alpha}$ or a thermal gradient $\delta f^{T\alpha}$ in direction $\alpha$, defined as $\delta f_{\alpha \boldsymbol{k}b} = \delta f^{E\alpha} E^{\alpha}$ and $\delta f_{\alpha \boldsymbol{k}b} = \delta f^{T\alpha} \nabla_{\alpha} T$ respectively.
These perturbations induce both a charge flux $\boldsymbol{J}$ and an energy flux $\boldsymbol{Q}$ defined as
\begin{equation}
\boldsymbol{J} = \frac{e}{N_k V} \sum_{\boldsymbol{k}b} v_{\alpha\boldsymbol{k}b}  \delta f_{\boldsymbol{k}b} \;,
\end{equation}
\begin{equation}
\boldsymbol{Q} = \frac{1}{N_k V} \sum_{\boldsymbol{k}b} v_{\alpha\boldsymbol{k}b} (\epsilon_{\boldsymbol{k}b}-\mu) \delta f_{\boldsymbol{k}b} \;.
\end{equation}
These fluxes can be compared with the Onsager relations
\begin{equation}
\boldsymbol{J} = L_{EE} \boldsymbol{E} + L_{ET} \nabla T \;,
\end{equation}
\begin{equation}
\boldsymbol{Q} = L_{TE} \boldsymbol{E} + L_{TT} \nabla T \;,
\end{equation}
from which we can define and compute the electronic transport coefficients, such as the electrical conductivity $\sigma$
\begin{equation}
\label{def:sigma}
    \sigma = L_{EE} \;,
\end{equation}
the drift mobility $\mu=\frac{\sigma}{ed}$, where $d$ is the carrier concentration, the Seebeck coefficient $S$
\begin{equation}
\label{def:seebeck}
    S = -L^{-1}_{EE} L_{ET} \;,
\end{equation}
and electronic thermal conductivity tensor $\kappa_{el}$
\begin{equation}
\label{def:kappa}
    \kappa_{el} = L_{TT} - L_{TE} L^{-1}_{EE} L_{ET} \;.
\end{equation}
With these definitions, we are finally ready to predict the thermal and electronic transport properties of crystals.


\subsection{BTE Solvers}
\label{Section:BteSolvers}
The similarity between electron and phonon BTEs allows us to readily implement solvers which can be applied generically to either the electron or phonon cases.
We first look for a solution that is linear in the external perturbation of the system, i.e. an out-of-equilibrium population in the form $f_{\lambda} \to f_{\lambda} E$ or $f_{\lambda} \to f_{\lambda} \nabla T$, with $\lambda$ as a combined index labeling the relevant band and momentum state (phonon or electron).
Under this linear approximation, we write the BTE (Eqs. (\ref{phBte}) (\ref{elBte1}) and (\ref{elBte2})) in the more concise form of a linear algebra problem
\begin{equation}
    \sum_{\lambda'} \Omega_{\lambda \lambda'} f_{\lambda'} = b_{\lambda} \;,
\end{equation}
where $b$ is the diffusion term, $\Omega$ is the scattering matrix, all chosen appropriately according to the transport problem under consideration.
Note that for the phonon transport problem, Phoebe uses at times a symmetrized scattering matrix $\tilde{\Omega}$ in place of $\Omega$ (see Ref. \cite{relaxons}) or the scattering matrix $A$ (see Ref. \cite{fugalloVariational}).

On the surface, the BTE appears as a trivial linear algebra problem; however, there are several complexities which arise when solving the BTE.
First, the scattering matrix as in Eqs. (\ref{elScatteringMatrix}) and (\ref{phScatteringMatrix}) cannot be directly inverted to solve the problem \cite{PhysRev.148.766} due to the presence of zero eigenvalues (arising from energy conservation and/or charge conservation). 

The second challenge is the sheer size of the problem: in order to obtain converged results, the summation over the Brillouin zone ($\lambda$ includes a wavevector index) must be performed for an extremely fine mesh of wavevectors. It is often necessary to use more than $10^5$ wavevectors.
Therefore, Phoebe contains a number of different solvers which operate at various levels of complexity and computational cost, each with distinct advantages and disadvantages, and lets the user decide which one is best suited for a particular material. 
Taking advantage of the similarity between charge and heat transport, the solvers have been implemented for both the electron and phonon BTE.


\subsubsection{Constant Relaxation Time Approximation}
The constant relaxation time approximation (CRTA) is a rudimentary, approximate solution to the BTE.
In this case, the scattering matrix is replaced with a single scalar
\begin{equation}
\Omega_{\lambda \lambda'} 
\approx 
\frac{1}{\tau} \;,
\end{equation}
where $\tau$ is a user-specified constant relaxation time.
Most of the complexity of the BTE is removed and the BTE is readily solved with the out-of-equilibrium population $f_{\lambda'} \approx \tau b_{\lambda}$.
The CRTA may provide reasonable results for the Seebeck coefficient, as this often has a weak dependence on the scattering rate, and may be used for fast estimates of trends in other transport properties.
However, the choice of $\tau$ is largely arbitrary and in many crystals a more detailed description of scattering processes must be included to quantitatively predict transport properties or even capture qualitative transport anomalies \cite{fedorova2021anomalous,kozinsky2021thermoelectrics}.


\subsubsection{Relaxation Time Approximation}
\label{Section:rtaSolver}
The relaxation time approximation (RTA) is another approximate solution to the BTE, in which only the diagonal elements of the scattering matrix are retained so that
\begin{equation}
\Omega_{\lambda \lambda'} 
\approx \Omega_{\lambda \lambda'} \delta_{\lambda \lambda'} =
\delta_{\lambda \lambda'} \frac{1}{\tau_{\lambda}} \;,
\end{equation}
where $\tau_{\lambda}$ is a carrier's lifetime, and the off-diagonal matrix elements are neglected.
With this approximation, the out-of-equilibrium population is readily found as $f_{\lambda'} \approx b_{\lambda} \tau_{\lambda}$.
The RTA is a fast and memory-efficient BTE solver, and can be used when more accurate solvers become prohibitively expensive.
For electronic transport, it often shows good agreement with more accurate solvers.
On the other hand, there are reported cases for which the RTA approximation has failed, particularly for materials where the conservation of crystal momentum is of the utmost relevance, such as in hydrodynamic phonon transport regimes \cite{nature-mio, fugallo-nano, PhysRevB.82.115427}. 
In these cases, one must use a solver which utilizes the complete scattering matrix, such as the solvers listed below. 


\subsubsection{Iterative solver}
\label{Section:IterativeSolver}
The iterative BTE solver \cite{sparavigna:nc, sparavigna:prb53, PhysRevB.72.014308} constructs the out-of-equilibrium populations with a geometric series.
First, we decompose the scattering matrix into its diagonal terms $\Omega^{out}$ and off-diagonal components $\Omega^{in}$.
A solution is found by iteratively correcting the RTA solution, $f^{\mathrm{RTA}} = \frac{1}{\Omega^{out}}b = \tau b$, as
\begin{equation}
f^i 
= 
\sum_{j=0}^i
\bigg(
-\frac{1}{\Omega^{out}} \Omega^{in}
\bigg)^j
\frac{1}{\Omega^{out}} b \;,
\end{equation}
where $j$ is an iteration index that can be stopped when transport coefficients are converged.
Note also that because $\Omega^{out}$ is a diagonal matrix, its inverse is trivially constructed.

Additionally, we implemented crystal symmetries in this solver to further accelerate simulations (see Section \ref{Section:Symmetries} below).
However, it's worth pointing out that the iterative solver can only converge as a geometric series, i.e. only if the absolute value of the determinant $|\det(\frac{1}{\Omega^{out}} \Omega^{in})|$ is smaller than 1.
If the convergence criterion is not satisfied, the algorithm may diverge.
The iterative method can be used both by opting to store the scattering matrix in memory, which allows a fast solution once the scattering matrix is built but at the expense of a large usage of memory, or not storing $\Omega$ in memory, so that we lower memory requirements but require to compute the action of the scattering matrix $\Omega\cdot f$ several times.


\subsubsection{Variational solver}
The variational BTE solver \cite{PhysRevB.88.045430, bonini2016} operates by minimizing the following quadratic functional of the carrier population $f$
\begin{equation}
\mathcal{F}[f] 
= 
\frac{1}{2}
f \cdot A \cdot f - b \cdot f \;.
\end{equation}
Since the scattering matrix is semi-positive definite \cite{hardy-ss, PhysRevB.88.045430, bonini2016} (a property related to the ever increasing value of entropy \cite{ziman}), the functional $\mathcal{F}$ has a single minimum which coincides with the solution of the BTE.
We implemented a conjugate gradient algorithm \cite{PhysRevB.88.045430} that minimizes the functional $\mathcal{F}$ and finds the exact solution to the BTE.
In contrast to the iterative solver, the variational solver is guaranteed to converge to the solution provided that the scattering matrix is semi-positive definite.
Like the iterative solver, the user can choose to store the matrix in memory to quickly build the solution, or not, so that the solution is found more slowly but with smaller memory requirements.


\subsubsection{Relaxons solver}
The last solver is based on a direct diagonalization of the BTE \cite{relaxons, PhysRevLettChaput}.
The scattering matrix is diagonalized with a ScaLAPACK solver, and we compute the complete set of eigenvalues and eigenvectors
\begin{equation}
\sum_{\lambda'} \Omega_{\lambda \lambda'} \theta_{\lambda'}^{\alpha}
= 
\frac{1}{\tau_{\alpha}} \theta_{\lambda}^{\alpha} \;,
\end{equation}
where $\alpha$ is an eigenvalue index, $\tau_{\alpha}$ is the inverse eigenvalue and $\theta_{\lambda}^{\alpha}$ the eigenvector.
Thanks to the completeness of the basis set of eigenvectors, we expand the out-of-equilibrium population as
\begin{equation}
\label{eq:relaxons}
f_{\lambda} = \sum_{\alpha} f_{\alpha} \theta_{\lambda}^{\alpha} \;,
\end{equation}
where we introduced auxiliary coefficients $f_\alpha$.
Next, we write the BTE in the eigenvector (relaxons) basis as
\begin{equation}
\sum_{\lambda} b_{\lambda} \theta_{\lambda}^{\alpha}
= 
\frac{1}{\tau_{\alpha}} f_{\alpha}  \;,
\end{equation}
and note that in this form we can analytically solve the BTE by substituting $f_{\alpha} = \tau_{\alpha} \sum_{\lambda} b_{\lambda} \theta_{\lambda}^{\alpha}$ in Eq. (\ref{eq:relaxons}).

This last solver has the benefit of always providing the exact solution, as it's not based on any iterative scheme and therefore has none of the related numerical instabilities.
Furthermore, this analytical solution to the BTE allows for further physical insights, such as the calculation of viscosity \cite{PhysRevX.10.011019} (see Section \ref{section:viscosity}).
On the other hand, this solver has the longest time-to-solution for calculations of transport properties.


\subsection{Electron-Phonon Averaged approximation}
The Electron Phonon Averaged (EPA) approximation offers another approximate solution to the BTE using an average over the phonon energies and coarse-mesh electron-phonon matrix elements \cite{SamsonidzeEPA}.
Although the Wannier function technique discussed in Section \ref{Section:ElectronProperties} provides an accurate interpolation of the electronic properties, the number of fine-mesh momentum vectors needed to converge transport calculations can be very large. Furthermore, the construction of the MLWF is a challenging task, not yet fully automated (progress is being made in this direction \cite{automaticWannier}), and therefore is not always suitable for materials studies.
Avoiding momentum space integration and Wannier interpolation altogether, EPA provides a way to estimate electron transport properties in a much faster and automatable way, suitable for rapid materials screening applications.

Consider a set of electronic energies $\epsilon (\boldsymbol{k}_{co})$ (at fixed band index) computed by a first-principles code on a coarse grid of wavevectors in the BZ.
To interpolate the energies, we implemented a plane-wave based interpolation method \cite{PhysRevB.38.2721}, which fits an electronic band with the periodic function
\begin{equation}
    \tilde{\epsilon} (\boldsymbol{k})
    =
    \sum_{m=0}^{M-1} c_m e^{i\boldsymbol{k}\cdot \boldsymbol{R}_m} \;,
\end{equation}
where $\boldsymbol{R}_m$ is a set of $m=0,1,\dots,M-1$ Bravais lattice vectors chosen as all $\boldsymbol{R}_m$ such that $|\boldsymbol{R}_m| < R_{cut}$, where $R_{cut}$ is a user-specified parameter, and $c_m$ is an expansion coefficient.
The expansion coefficients are found by optimizing a target error function against a set of training points.
Further details can be found in Ref. \cite{PhysRevB.38.2721}.
This technique doesn't involve Wannier functions and is thus simpler to implement and automate.
In practice, compared to the Wannier interpolation, the sampling of wavevectors $\{\boldsymbol{k}_{co}\}$ typically needs to be denser in order to converge transport results.
With this interpolation procedure, it is possible to evaluate the electronic density of states $\rho(\epsilon) = \frac{1}{V N_k} \sum_{kb} \delta(\epsilon - \tilde{\epsilon}_b(\boldsymbol{k}) )$, where $\tilde{\epsilon}_b(\boldsymbol{k})$ are the interpolated energies of each band $b$.

The next component of the EPA approach is a scheme for estimating the electron-phonon coupling.
The idea underlying the EPA method is to replace momentum-space integrations of the BZ with integrations over the quasiparticle energies, simplifying some of the microscopic details of electronic transport to gain computational performance.
The electron-phonon coupling is approximated as
\begin{equation}
    |g^2_{bb'\nu}(\boldsymbol{k},\boldsymbol{q})|^2
    \to
    g^2_{\nu}(\epsilon_1,\epsilon_2) \;,
\end{equation}
which replaces a 6-dimensional wavevector dependence on $\boldsymbol{k}$ and $\boldsymbol{q}$ with a 2-dimensional dependence on electronic energies.
Additionally, as $g$ often has a weak energy dependence, it is possible to sample $g^2$ with just a few energy values.
There are several ways to construct the function $g^2$ \cite{SamsonidzeEPA}, and we implemented the moving-least-squares (MLS) method \cite{BANG201822}, which constructs $g^2$ by minimizing the integral
\begin{equation}
\sum_{bb'}
\int \mathrm{d} \boldsymbol{k}
\int \mathrm{d} \boldsymbol{q}
\left(|g_{bb'\nu}(\boldsymbol{k},\boldsymbol{q})|^2
    -
    g^2_{\nu}(\epsilon_1,\epsilon_2)\right)^2
    \exp \bigg( -\frac{(\epsilon_{\boldsymbol{k}b}-\epsilon_1)^2+(\epsilon_{\boldsymbol{k}b'}-\epsilon_2)^2}{2\sigma^2_{\text{gauss}}} \bigg) \;,
\end{equation}
where $\sigma_{\text{gauss}}$ is a user-specified smoothing scale parameter.
The energy dependence on $\epsilon_1$ and $\epsilon_2$ is implemented numerically with a user-specified uniform sampling of an energy interval.

Once the coupling is constructed, the energy-dependent electronic lifetime is calculated as
\begin{align}
    \tau^{-1}(\epsilon,\mu,T)
    =
    \frac{2\pi V}{g_s \hbar}
    \sum_{\nu}
    \{
    & g^2_{\nu}(\epsilon,\epsilon+\bar{\omega}_\nu))^2
    \left[n(\bar{\omega}_\nu,T)+f(\epsilon+\bar{\omega}_\nu,\mu,T)\right] \rho(\epsilon+\bar{\omega}_\nu) \\
    &+
    g^2_{\nu}(\epsilon,\epsilon-\bar{\omega}_\nu))^2
    \left[n(\bar{\omega}_\nu,T)+1-f(\epsilon-\bar{\omega}_\mu,\mu,T)\right] \rho(\epsilon-\bar{\omega}_\nu)
    \}  \;, \nonumber
\end{align}
where $\bar{\omega}_\nu = \frac{1}{N_q} \sum_{\boldsymbol{q}} \omega_{\boldsymbol{q}\nu}$ is the average phonon frequency of the phonon branch $\nu$, $f$ and $n$ are energy-dependent Fermi--Dirac and Bose--Einstein distributions respectively, and $g_s$ is the spin-multiplicity (equal to 2 for non spin-polarized calculations).

Using this energy-dependent lifetime, the BTE is readily solved at the RTA level, and it is possible to evaluate the electronic transport coefficients using an energy-dependent approximation for Eqs. (\ref{def:sigma}), (\ref{def:seebeck}), (\ref{def:kappa}), as discussed in greater details in Ref. \cite{SamsonidzeEPA}.
Typically, the computational bottleneck for this method in Phoebe is the construction of the electronic density of states, which simply has a cost linear in the number of wavevectors used to sample the BZ.


\subsection{Viscosity}
\label{section:viscosity}
In some materials with large conductivities, transport doesn't follow the conventional macroscopic diffusion equations such as Fourier's law for heat or Ohm's law for charge currents.
Recent studies highlighted the necessity to complement conventional transport coefficients such as conductivity with hydrodynamic transport coefficients such as viscosity.
Following Ref. \cite{PhysRevX.10.011019}, we implemented the phonon viscosity as defined from the response coefficient $\eta$,
\begin{equation}
    \Pi^{ij} = \eta^{ijkl} \frac{\partial u^k}{\partial r^l} \;,
\end{equation}
where $i, j, k, l$ are Cartesian indices, $u$ is the phonon drift velocity in direction $k$, $\boldsymbol{r}$ is the position, and $\Pi$ is
the crystal momentum flux $\Pi^{ij} = \frac{1}{V N_{\boldsymbol{q}}} \sum_{\boldsymbol{q}\nu} \hbar q^i v^j \delta n^{D}_{\boldsymbol{q}\nu}$, where $\delta n^{D}_{\boldsymbol{q}\nu}$ is the out-of-equilibrium phonon population in response to a perturbation of the drift velocity $\boldsymbol{u}$.
The out-of-equilibrium population can be found solving a modified version of the BTE, as reported in full details in Ref. \cite{PhysRevX.10.011019}.
Finally, the phonon viscosity $\mu$ is found by symmetrizing $\eta$ as
\begin{equation}
    \mu^{ijkl} = \frac{\eta^{ijkl}+\eta^{ilkj}}{2} \;.
\end{equation}
The phonon viscosity can be used to parametrize a generalization of the Fourier's law for describing heat transport.
We also implemented the analogous expression for electronic viscosity, which is readily obtained by replacing phonon properties such as group velocity, equilibrium distribution and lifetimes with their electronic counterparts.


\subsection{Transport with the Wigner distribution}
In systems where carriers' energies overlap with each other, such as in strongly anharmonic systems \cite{simoncelli2019} or narrow-gap semiconductors \cite{CEPELLOTTI2021100412}, it is important to include contributions to transport from the coupling of multiple bands, an effect which is not described by the BTE.
To accurately describe these contributions, we implemented the Wigner transport corrections for phonon and electron transport properties. 
The starting point of this modeling is the Wigner distribution function of phonons $N_{\nu\nu'}(\boldsymbol{r},\boldsymbol{q},t)$ or electrons $F_{bb'}(\boldsymbol{r},\boldsymbol{k},t)$.
The Wigner distribution cannot be interpreted anymore as a probability distribution, as is the case for electron and phonon population appearing in the BTE.
Instead, $N$ and $F$ are quasiprobability distributions that describe the state of a quantum system.

It is possible to construct an approximate equation of motion for the Wigner distribution function. As in the spirit of the BTE, this expression aims to describe the out-of-equilibrium state of a quantum system.
In particular, the equation of motion for the phonon Wigner distribution function $N$ is
\begin{equation}
    \frac{\partial N(\boldsymbol{r},\boldsymbol{q},t)}{\partial t}
    + i [\tilde{\Omega}(\boldsymbol{q}),N(\boldsymbol{r},\boldsymbol{q},t)]
    +\frac{1}{2} \{\boldsymbol{V}(\boldsymbol{q}),\cdot \nabla_{\boldsymbol{r}}N(\boldsymbol{r},\boldsymbol{q},t)
    \}
    =
    \frac{\partial N(\boldsymbol{r},\boldsymbol{q},t)}{\partial t} \Bigg|_{\mathrm{coll}} \;,
\end{equation}
where $\frac{\partial N(\boldsymbol{r},\boldsymbol{k},t)}{\partial t} \Big|_{\mathrm{coll}}$ is the phonon scattering operator, $\tilde{\Omega}(\boldsymbol{q})$ is a matrix of energies with elements $\tilde{\Omega}_{\nu\nu'}(\boldsymbol{q}) = \delta_{\nu\nu'} \hbar \omega_{\boldsymbol{q}\nu}$, $[,]$ is a commutator and $\{,\}$ an anticommutator.
The electronic version of the Wigner transport equation is
\begin{equation}
    \frac{\partial F(\boldsymbol{r},\boldsymbol{k},t)}{\partial t} 
    + i \left[\mathcal{E}(\boldsymbol{k})+\boldsymbol{d}(\boldsymbol{k})\cdot \boldsymbol{E},F(\boldsymbol{r},\boldsymbol{k},t)\right]
    +\frac{1}{2} \{\boldsymbol{V}(\boldsymbol{k}),\cdot \nabla_{\boldsymbol{r}} F(\boldsymbol{r},\boldsymbol{k},t) \}
    -e\boldsymbol{E} \cdot \nabla_{\boldsymbol{k}} F(\boldsymbol{r},\boldsymbol{k},t)
    =
    \frac{\partial F(\boldsymbol{r},\boldsymbol{k},t)}{\partial t} \Bigg|_{\mathrm{coll}} \;,
\end{equation}
where $\frac{\partial F(\boldsymbol{r},\boldsymbol{k},t)}{\partial t} \Big|_{\mathrm{coll}}$ is the electron scattering operator, $\mathcal{E}(\boldsymbol{k})$ is a matrix of energies with elements $\mathcal{E}_{bb'}(\boldsymbol{k}) = \delta_{bb'} \epsilon_{\boldsymbol{k}b}$, and $\boldsymbol{d}_{bb'}(\boldsymbol{k})$ is the electronic dipole \cite{CEPELLOTTI2021100412}.
The scattering operator for electrons is
\begin{equation}
    \frac{\partial F(\boldsymbol{r},\boldsymbol{k},t)}{\partial t} \Bigg|_{\mathrm{coll}} 
    = 
    - \sum_{\boldsymbol{k}'b'} \Omega_{\boldsymbol{k}b,\boldsymbol{k}'b'} F_{b'b'}(\boldsymbol{r},\boldsymbol{k},t)
    -(1-\delta_{bb'}) \frac{ \tau^{-1}_{\boldsymbol{k}b} + \tau^{-1}_{\boldsymbol{k}b'} }{2} F_{bb'}(\boldsymbol{r},\boldsymbol{k},t)
     \;,
\end{equation}
where we distinguished the scattering matrix acting on the diagonal components of the Wigner distribution matrix from the off-diagonal terms of the scattering operator.
The scattering operator for the phonon Wigner distribution is obtained in full analogy by replacing in the previous equation electron lifetimes and scattering matrix with the phonon counterparts.
The details of the derivation of these equations are given in Refs. \cite{simoncelli2019, CEPELLOTTI2021100412}.

There are both similarities and notable differences between these equations and the electron or phonon BTE.
First of all, the BTE is obtained from the Wigner transport equation as the limit in which the off-diagonal components of the Wigner distributions are neglected (i.e. $N_{\nu\nu'}\approx N_{\nu\nu'}\delta_{\nu\nu'}$ or $F_{bb'}\approx F_{bb'}\delta_{bb'}$).
As a result, the Wigner formalism expands on the well-established predictive capability of the BTE.
Critically, the equation of motion for the Wigner distribution adds a commutator, which allows coupling between different off-diagonal elements of the Wigner distribution function.

We note that the diagonal (single-band) elements of these equations coincide with the electron and phonon BTE discussed previously, and that results for transport coefficients can be interpreted as corrections to the BTE predictions.
Furthermore, these Wigner corrections to the BTE transport coefficients only depend on the carrier lifetimes. Because these quantities are already computed when constructing the scattering operator for the BTE, the computational cost of the Wigner corrections is negligible compared to the cost of constructing and solving the BTE.

Just as described for the BTE, the Wigner transport equations for electrons and phonons can be solved considering linear perturbations of the Wigner distribution from equilibrium.
Once the equations are solved, the transport coefficient are readily found.
For example, the expectation value of charge flux is
\begin{equation}
    J(\boldsymbol{r},t) = \frac{g_s}{2 V N_k} \sum_{\boldsymbol{k}b} \Big[ \{ \boldsymbol{V}(\boldsymbol{k}), F(\boldsymbol{r}, \boldsymbol{k}, t) \} \Big]_{bb} \;,
\end{equation}
where $g_s$ takes into account the spin degeneracy (here the equation is shown for a system without spin polarization).
As shown in Ref. \cite{CEPELLOTTI2021100412}, the electrical conductivity can be written as $\sigma = \sigma^{BTE} + \Delta \sigma$, where $\sigma^{BTE}$ is the conductivity predicted by the BTE, and $\Delta \sigma$ is a correction introduced by the Wigner distribution function:
\begin{equation}
    \Delta \sigma_{\alpha\beta}
    =
    \frac{2 g_s e^2}{V N_k} \sum_{\boldsymbol{k}bb'} 
    \frac{ \bar{f}_{b'}(\boldsymbol{k}) - \bar{f}_{b}(\boldsymbol{k}) }{ \epsilon_{b'}(\boldsymbol{k}) - \epsilon_{b}(\boldsymbol{k}) }
    v_{\alpha \boldsymbol{k}, bb'} v^{*}_{\beta \boldsymbol{k} b'b}
    \frac{ \tau^{-1}_{\boldsymbol{k}b} + \tau^{-1}_{\boldsymbol{k}b'} }{ 4(\epsilon_{b'}(\boldsymbol{k}) - \epsilon_{b}(\boldsymbol{k}))^2 + (\tau^{-1}_{\boldsymbol{k}b} + \tau^{-1}_{\boldsymbol{k}b'})^2 } \;.
\end{equation}
The correction to conductivity introduced by the Wigner distribution can be significant, for example when the difference between the energies of quasiparticles in valence and conduction bands are small, i.e. for narrow-gap semiconductors, or when the dipole coupling between valence and conduction states is strong.
Corrections to the other Onsager coefficients can similarly be derived.

A similar result holds for the phonon transport, where the energy current can be computed as
\begin{equation}
    \boldsymbol{Q}(\boldsymbol{r},t) = \frac{1}{2 V N_q} \sum_{\boldsymbol{q}\nu} \Big[ \{ \boldsymbol{V}(\boldsymbol{q}), N(\boldsymbol{r}, \boldsymbol{q}, t) \} \Omega(\boldsymbol{q}) \Big]_{\nu\nu} \;.
\end{equation}
Similarly, the lattice thermal conductivity is $k=k^{BTE}+\Delta k$, with the first term being the lattice thermal conductivity predicted by the phonon BTE, and the Wigner distribution resulting in the correction
\begin{equation}
    \Delta k_{\alpha\beta}
    =
    \frac{\hbar^2}{k_B T^2 V N_q} 
    \sum_{\boldsymbol{q}\nu\nu'} 
    \frac{ \omega_{\boldsymbol{q}\nu} + \omega_{\boldsymbol{q}\nu'} }{ 2 }
    v_{\alpha \boldsymbol{q}, \nu\nu'} v^{*}_{\beta \boldsymbol{q} \nu'\nu}
    (\omega_{\boldsymbol{q}\nu}\bar{n}_{\boldsymbol{q}\nu} (\bar{n}_{\boldsymbol{q}\nu}+1) + \omega_{\boldsymbol{q}\nu'}\bar{n}_{\boldsymbol{q}\nu'} (\bar{n}_{\boldsymbol{q}\nu'}+1) )
    \frac{ \tau^{-1}_{\boldsymbol{q}\nu} + \tau^{-1}_{\boldsymbol{q}\nu'} }{ 4(\omega_{\boldsymbol{q}\nu'} - \omega_{\boldsymbol{q}\nu})^2 + (\tau^{-1}_{\boldsymbol{q}\nu} + \tau^{-1}_{\boldsymbol{q}\nu'})^2 } \;.
\end{equation}
It has been shown \cite{simoncelli2019} that this correction is especially relevant for crystals with localized vibrational modes, when the small phonon group velocities suppresses the BTE lattice thermal conductivity and small lifetimes and small energy differences between phonon modes enhance the correction described by the Wigner distribution.


\subsection{Dirac delta and Smearing}
\label{dirac_delta_section}
Calculation of materials properties such as scattering rates often requires a numerical approximation of the Dirac--delta function, for example, the density of states, $\rho$
\begin{equation}
    \rho(\epsilon) = \frac{1}{2\pi^3 V} \sum_b \int \delta(\epsilon-\epsilon_{\boldsymbol{k}b})  \mathrm{d} \boldsymbol{k} \;.
\end{equation}
We offer three different schemes for the numerical approximation of the Dirac--delta for the user to choose from: 
\begin{enumerate}
    \item \emph{Gaussian smearing}: the Dirac--delta is replaced with a Gaussian function 
    \begin{equation}
    \delta(\epsilon_{kb}-\epsilon_{k'b'})
    \approx
    \frac{1}{\sqrt{\pi} \sigma}
    exp\left(- \frac{(\epsilon_{\boldsymbol{k}b}-\epsilon_{k'b'})^2}{\sigma^2} \right) \;,
    \end{equation}
    where $\sigma$ is a user-specified smearing parameter.
    This method is simple and fast, but requires the user to converge results with respect to $\sigma$ in conjunction with the number of wavevectors used for the BZ integration.
    
    \item \emph{Adaptive Gaussian smearing}: the Gaussian smearing parameter $\sigma$ introduced above can be approximated \cite{adaptiveGaussian1,adaptiveGaussian2} as
    \begin{equation}
    \sigma = 
    \frac{A}{\sqrt{12}}
    \sqrt{ \sum_\beta \left(\sum_{\alpha} \left(v^{\alpha}_{\boldsymbol{k}b}-v^{\alpha}_{\boldsymbol{k}'b'}\right) \frac{C_{\alpha\beta}}{N_{\beta}} \right) } \;,
    \end{equation}
    where $C$ is the matrix describing the crystal lattice vectors, $N_\beta$ is the size of the Monkhorst-Pack mesh in crystal direction $\beta$, and the parameter $A$ is a prefactor that can be fixed to 1 \cite{adaptiveGaussian2}.
    As this method self-adjusts $\sigma$ with respect to the chosen BZ sampling and the local derivative of the carrier's energy, it tends to yield easier convergence of results compared the fixed-width Gaussian smearing scheme.
    
    \item \emph{Tetrahedron method}: discussed in Ref. \cite{tetrahedronMethod}, this method provides a truly parameter-free linear approximation of the integrand. 
    For simplicity, the tetrahedron method is currently implemented only for the calculation of the DoS. 
    The method is accurate and parameter-free, although considerably slower than the above two alternatives.
\end{enumerate}


\subsection{Symmetries}
\label{Section:Symmetries}
A crystal is characterized by a set of symmetry operations $S=\{R,t\}$, with each operation $S$ consisting of a rotation $R$ and a fractional translation $t$ such that the application of a translation and a rotation to the atomic positions leaves the overall crystal unchanged.
In reciprocal space, instead of working with a set of wavevectors $\{\boldsymbol{k}\}$ that uniformly samples the BZ, symmetries allow the selection of a subset of irreducible wavevectors $\{\boldsymbol{k}^*\}$, such that the complete set of wavevectors can be obtained by applying rotations to the irreducible wavevectors set $\boldsymbol{k} = R \boldsymbol{k}^*$.
As a result, integration over the BZ can be sped up by restricting the domain to the irreducible wedge of the BZ.

We recall here that microscopic carrier properties also obey symmetry relations.
For example, scalars such as the carriers energy or lifetimes are left invariant by symmetry operations between an irreducible point $\boldsymbol{k}^*$ and its symmetry equivalent point $\boldsymbol{k} = R\boldsymbol{k}^*$ \cite{PhysRevLettChaput}
\begin{equation}
    \epsilon_{\boldsymbol{k} b} 
    =
    \epsilon_{\boldsymbol{k}^* b} \;, 
\end{equation}
and
\begin{equation}
    \tau_{\boldsymbol{k} b} 
    =
    \tau_{\boldsymbol{k}^* b} \;.
\end{equation}
Instead, the group velocity transforms like a vector as
\begin{equation}
    v_{\alpha \boldsymbol{k} b} 
    =
    \sum_{\beta} R_{\alpha\beta} 
    v_{\beta \boldsymbol{k}^* b} \;,
\end{equation}
and similarly, the out-of-equilibrium population \cite{PhysRevLettChaput} transform as
\begin{equation}
    \delta f_{\alpha \boldsymbol{k} b} 
    = \sum_{\beta} R_{\alpha\beta} 
    \delta f_{\beta \boldsymbol{k}^* b} \;.
\end{equation}
As a result, the calculation of several properties can be sped up by restricting states to the irreducible wedge of the BZ, for example when computing the density of states, or solving the BTE at the RTA level.

The relaxation time approximation and iterative solvers (see Sections \ref{Section:rtaSolver} and \ref{Section:IterativeSolver}) support the use of symmetries. The BTE for an irreducible point $\boldsymbol{k}^*$ (omitting band indices for simplicity) is \begin{equation}
    b_{\boldsymbol{k}^*}^{\alpha} 
    = \sum_{\boldsymbol{k}'} A_{\boldsymbol{k}^*\boldsymbol{k}'} f_{\boldsymbol{k}'}^{\alpha}
    = \sum_{\beta \boldsymbol{k}'} A_{\boldsymbol{k}^*\boldsymbol{k}'} R^{\alpha\beta} f_{\boldsymbol{k}'}^{\beta}
    = \sum_{\beta \boldsymbol{k}^{'*}} \tilde{A}_{\alpha\beta;\boldsymbol{k}^{*}\boldsymbol{k}^{'*}} f_{\boldsymbol{k}^{'*}}^{\beta} \;,
\end{equation}
where we used the symmetry properties of the out-of-equilibrium population, and $\tilde{A}$ is a tensor describing the scattering properties limited to the irreducible BZ wedge.
As a result of symmetries, the BTE's linear algebra problem can be solved with a much smaller number of momentum states, and thus with a better computational efficiency.


\section{Software}

Phoebe is written in \verb_C++_ following the object-oriented paradigm.
Since Phoebe computes a range of extremely different physical properties, each of these properties are organized in Apps.
A main function simply initializes global properties (such as starting the MPI or Kokkos environments), reads the user input and launches the appropriate App through a class factory.
Currently, Phoebe offers the following Apps:
\begin{itemize}
    \item \verb+ElectronWannierBandsApp+, \verb+ElectronFourierBandsApp+, \verb+PhononBandsApp+: calculation of electron or phonon energies along a BZ path;
    \item \verb+ElectronWannierDosApp+, \verb+ElectronFourierDosApp+, \verb+PhononDosApp+: calculation of electron or phonon DOS;
    \item \verb+ElPhQeToPhoebeApp+: processing of the electron-phonon coupling from QE to a Phoebe format. 
    An EPA processing computes the electron-phonon averaged coupling, while the Wannier processing transforms the coupling to its real-space Wannier representation;
    \item \verb+ElectronLifetimesApp+ and \verb+PhononLifetimesApp+: calculation of carriers' lifetimes along a BZ path;
    \item \verb+ElectronWannierTransportApp+: calculation of electronic transport coefficients with Wannier interpolation method;
    \item \verb+TransportEpaApp+: calculation of electronic transport coefficients with the EPA approximation;
    \item \verb+PhononTransportApp+: evaluation of phonon transport properties.
\end{itemize}
The physical properties computed by these Apps are either printed to standard output or written to JSON files for user's convenience.
We also provide a few \verb+Python+ scripts to generate some of the most common plots of physical properties.

Each App uses a number of classes built to reduce code duplication and to provide convenient data structures.
In fact, a central goal of the project is to guarantee a simple deployment of methods and algorithms for future releases.
The most important among the several classes available are
\begin{itemize}
    \item \verb+Crystal+: description of the crystal unit cell and crystal symmetries;
    \item \verb+Points+: handling of wavevectors in the BZ;
    \item \verb+HarmonicHamiltonian+: base class for the calculation of non-interacting carriers properties; subclassed into \verb+PhononH0+, \verb+ElectronH0Wannier+ and \verb+ElectronH0Fourier+ for evaluating phonon properties, Wannier- and Fourier-interpolated electron properties respectively;
    \item \verb+BandStructure+: storage of carriers' energies, velocities and eigenvectors;
    \item \verb+Context+: description of user input variables;
    \item \verb+MPIcontroller+: wrapper for calls to MPI functions;
    \item \verb+Matrix+: storage of MPI-distributed matrices and a wrapper to ScaLAPACK subroutines;
    \item \verb+StatisticsSweep+: calculation and description of values for temperature, chemical potentials and carrier concentrations;
    \item \verb+Interaction+: base class for the evaluation of the scattering coupling strength. Its subclasses describe the three-phonon interaction, electron-phonon coupling with Wannier interpolation and the EPA coupling;
    \item \verb+ScatteringMatrix+: base class for the calculation, storage and manipulation of the electron or phonon scattering matrix;
    \item \verb+VectorBTE+: data structure for storage and manipulation of the out-of-equilibrium population;
    \item \verb+Observable+: base class of a number of classes for evaluating transport coefficients.
\end{itemize}
In the following we will discuss some of these classes in greater detail.


\subsection{Points class}
The \verb+Points+ class in Phoebe manages the wavevectors used to sample the BZ.
We defined two constructors, one for instantiating a class for uniform sampling of the BZ and another for paths of wavevectors in the BZ.
When uniformly sampling the BZ with a Monkhorst-Pack grid, the complete list of points (wavevectors) is not explicitly stored for memory management convenience.
This implicit list contains points sorted by their coordinates (incrementing the coordinates z, y and x in this order).
When describing BZ paths, instead, points are not sorted, and their coordinates are explicitly stored in a list.
Points are internally manipulated in crystal coordinates (i.e. in scaled lattice vectors units), and we provide options to output points in Cartesian coordinates or to fold them in the first BZ.

In the \verb+Points+ class we implemented the logic for reducing a set of wavevectors to the set of symmetry-irreducible wavevectors in the BZ.
The class method \verb+setIrreduciblePoints()+, takes the set of crystal symmetries (from the \verb+Crystal+ class) and finds the list of irreducible BZ points, i.e. finds the irreducible set of points $\boldsymbol{k}^*$ such that all remaining points $\boldsymbol{k}$ can be found applying a rotation $R$ to $\boldsymbol{k}^*$.

It's worth noting that symmetry operations are not unique, so that a symmetry operation may rotate a wavevector correctly as $\boldsymbol{k}=R\boldsymbol{k}^*$ but other quantities such as the group velocity may not be rotated correctly $\boldsymbol{v}(\boldsymbol{k})\neq R \boldsymbol{v}(\boldsymbol{k}^*)$.
Since the carrier populations transform like group velocities, the method \verb+setIrreduciblePoints()+ can optionally find a set of symmetries that rotate both points and velocities.

The \verb+Points+ class also has a \verb+setActiveLayer()+ method that allows discarding some points that don't provide a sizeable contribution to a physical observable, a functionality that will be explained in Section \ref{Section:BandStructure} in conjunction with the \verb+ActiveBandStructure+ class.
After a call to this method, the functionalities in this class operate as if only the restricted set of points are present in the class.

Finally, several other parts of the code require the lookup of a wavevector in the list of points and, due to the large number of points used to sample the BZ (often in the excess of 10$^4$), this search operation can take a substantial part of the calculation if not implemented efficiently.
If \verb+Points+ describes a uniform BZ sampling with a complete Monkhorst-Pack grid, the point search in the list can be done in $o(0)$ operations, taking advantage of the implicit ordering of points.
If the list of points is sorted (e.g. after a call to \verb+setActiveLayer()+) the lookup operation can be done with a binary search algorithm, which scales logarithmically with the number of points.
Only if points are not sorted (e.g. when they describe a BZ path), the point search is made with a slow element-by-element comparison.


\subsection{Band structure classes}
\label{Section:BandStructure}
We developed a data structure dedicated to the storage of the band structure, that is, the energies ($\epsilon_{\boldsymbol{k}b}$ and $\hbar \omega_{\boldsymbol{q}\nu}$), eigenvectors ($U_{\boldsymbol{k}mb}$ and $z_{\boldsymbol{q}\nu}^{s\alpha}$) and velocities ($v_{\alpha\boldsymbol{k}bb'}$ and $v_{\alpha\boldsymbol{q}\nu\nu'}$) of carriers over the BZ.

To handle this information, we defined two classes: \verb+FullBandStructure+ and \verb+ActiveBandStructure+.
Both classes take a \verb+Points+ instance as input, since \verb+Points+ defines the set of wavevectors over which the carriers' properties are computed.
While both classes store the band structure, there is an important difference.
The \verb+FullBandStructure+ class stores energies, eigenvectors and velocities for all wavevectors defined in the \verb+Points+ instance.
For example, given $N_k$ points and $N_b$ electronic bands, \verb+FullBandStructure+ stores the complete list of $N_k \times N_b$ values of e.g. energies $\epsilon_{\boldsymbol{k}b}$ (and similar for eigenvectors or velocities).
Since wavevector meshes can be extremely large, carriers properties can optionally be stored in a MPI-distributed fashion, using the \verb+Matrix+ class described in Section \ref{Section:Matrix}.

\verb+ActiveBandStructure+ contains similar functionalities of \verb+FullBandStructure+ in that it stores information of carriers properties (in fact they share the same virtual parent class).
However, the constructor of this class discards some Bloch states $(\boldsymbol{k}b)$ that don't contribute significantly to the calculation of transport properties.
Because transport properties at finite temperature only depend on thermally excited electron and phonon states, there are many states which do not contribute and are readily discarded to save computational resources.
For example, only long-wavelength acoustic phonon states contribute to transport at low temperatures $T$ as their excitation energy goes to 0 with $T$.
Similarly, only electrons with energy close to the chemical potential $\mu$ contribute to charge currents and thus we can restrict calculations to electronic states such that $|\epsilon_{\boldsymbol{k}b}-\mu|$ is smaller than a few $k_BT$.
As a result, states that are not thermally excited can be discarded, and in Phoebe we term this band and momentum-space state filtering procedure as a "window".

We implemented two kinds of windows.
The first window type imposes an energy cutoff and discards all states with energy larger than a user-specified cutoff $\epsilon_{\mathrm{cutoff}}$: for phonons we retain states such that $|\hbar \omega_{\boldsymbol{q}\nu}|<\epsilon_{\mathrm{cutoff}}$ and for electrons $| \epsilon_{\boldsymbol{q}\nu}-\mu| < \epsilon_{\mathrm{cutoff}}$.
However, this kind of window is impractical, as the cutoff is temperature dependent and, in practice, a user would need to perform a convergence test on $\epsilon_{\mathrm{cutoff}}$.

We implemented a second window type, noting that transport coefficients contain terms of the form $\bar{f}_{\boldsymbol{k}b} (1- \bar{f}_{\boldsymbol{k}b})$ for electrons or $\bar{n}_{\boldsymbol{k}b} (1+\bar{n}_{\boldsymbol{k}b})$ for phonons (which arise from the energy or temperature derivative of equilibrium distributions).
For both carriers, these terms decay exponentially to zero if the carrier's energy is much larger than $k_BT$, thus suppressing contributions of high-energy carriers to transport properties.
We therefore define a "population window" by retaining only states such that, for electrons
\begin{equation}
    \bar{f}_{\boldsymbol{k}b} 
    (1- \bar{f}_{\boldsymbol{k}b}) > \delta \;,
\end{equation}
or for phonons
\begin{equation}
    \bar{n}_{\boldsymbol{k}b} 
    (1+\bar{n}_{\boldsymbol{k}b}) > \delta \;,
\end{equation}
where $\delta$ is a small number.
Since the temperature dependence is inside the definition of Bose--Einstein and Fermi--Dirac distributions, we found that $\delta$ can be hard-coded to a fixed value and that transport properties are not affected.
In this way, we achieve a reduction in the number of Bloch states used in the various calculations, and avoid the introduction of an additional parameter to be converged, for greater user convenience.

As a result of this filtering procedures, the number of electron bands $N_b$ or phonon branches $N_{\nu}$ is no longer a constant but depends on the wavevector index.
As a result, quantities like the electronic energy $\epsilon_{\boldsymbol{k}b}$ cannot be stored as a matrix of size $N_k \times N_b$. Instead, energies are stored as a flattened vector, and we provide class methods to access energies through combined band and wavevector state indices.
A similar strategy is adopted for storing eigenvectors or velocities.


\subsection{Matrix}
\label{Section:Matrix}
As described in Sections \ref{section:phBTE}, \ref{Section:ElectronBTE}, the BTE is a linear algebra problem and due to the number of Bloch states involved, requires distributed matrices and parallel linear algebra operations.
In order to perform these massively parallel dense linear algebra operations, we utilize the ScaLAPACK library \cite{scalapack}.
In order to both avoid explicit calls to this library in the main body of code and to allow future flexibility should we desire to swap ScaLAPACK routines with those of a different library, we developed the \verb+Matrix+ class as a wrapper to this library.

Because ScaLAPACK is mostly built in Fortran and C languages, it follows an imperative procedural programming paradigm, i.e. a set of routines that must be called in the right order, resulting in a departure from the object-oriented programming style adopted in the rest of Phoebe.
The \verb+Matrix+ class has therefore been designed to map this procedural-style library into an object-oriented style.
Therefore, the class constructor takes care of ScaLAPACK initializations, such as starting the BLACS environment and allocating the memory locations, while deallocations take place in the class destructor.
Various class methods have been implemented to provide wrappers around the needed linear algebra operations, such as matrix-matrix multiplication and matrix diagonalization, with the benefit of hiding within the class method all the necessary auxiliary calls (such as temporary memory allocations or index mappings) in order to call use the ScaLAPACK library.
As a result, the rest of Phoebe's code can act on distributed memory with a simplified interface that doesn't obfuscate the main body of the transport code.

\verb_C++_ class templates further allow us to generalize the class to store different data-types (in particular, real and complex floating point values).
We also allow the code to be compiled without the ScaLAPACK or MPI libraries. 
In this case, the \verb+Matrix+ class falls back to a "serial matrix" which relies on the \verb+Eigen+ library, losing all MPI-parallel characteristics (therefore becoming mostly suitable for debugging purposes).


\subsection{Scattering Matrix}
The \verb+ScatteringMatrix+ class, as the name suggest, implements operations related to BTE's scattering operator.
A key challenge in developing this class has been to try generalize this code to achieve different purposes.
In order to implement the calculation of scattering rates, we defined \verb+ScatteringMatrix+ as a virtual base class that defines a virtual method \verb+builder()+; the method is implemented in the subclasses \verb+PhononScatteringMatrix+ and \verb+ElectronScatteringMatrix+ and evaluates Eqs. (\ref{elScatteringMatrix}) or (\ref{phScatteringMatrix}).
In this way, auxiliary methods such as the calculation of a matrix-vector product or the scattering matrix diagonalization can be defined directly in the base class and are shared by the two different carriers.

The scattering matrix has a few modes of operation.
If the BTE is to be solved only at the RTA level, only the calculation of lifetimes is performed by the class instance.
For exact BTE solutions, for which the full scattering matrix must be dealt with, we distinguish two cases: first, when the complete scattering matrix is stored in memory or alternatively, when only the product of the scattering matrix on a population vector is computed (without the need of saving the matrix in memory).
In the former case, the calculation has a larger memory requirement, however, this allows to quickly solve the BTE. 
In the latter case, the memory footprint of the calculation is reduced, but as a drawback each iteration of a BTE solver will require more computational time.
We thus let the user choose the desired memory impact, which will depend on both the material studied and on the available hardware.
To summarize, the \verb+builder()+ method, depending on the input, is tasked with computing 1) the carriers' lifetimes, 2) the scattering matrix or 3) its action on a vector.
Note that due to the size of the scattering matrix $A_{\nu\nu'}$, it is stored as a class member using an MPI-distributed \verb+Matrix+ instance.

Great care has been taken to optimize the efficiency of the scattering rate calculations implemented in \verb+builder+, as these are some of the most computationally intensive functions in the code.
Significant improvements to the calculation speed have been achieved by precomputing all independent particle properties whenever possible.
Additionally, we used MPI parallelization to accelerate loops over $\boldsymbol{k}$ and $\boldsymbol{k}'$ wavevectors, while other loops over band indices are accelerated with OpenMP parallelization.
The calculation of the interaction coupling strength and its GPU acceleration is described in the next section.




\subsection{Kokkos (Interaction)}
\label{Section:Kokkos}

In Phoebe, great care has been taken to accelerate the calculation of interaction matrix elements and scattering rates. 
The calculation of the scattering coupling strength for either three phonon and electron-phonon scattering in reciprocal space, $f_{\alpha\beta\gamma}(\boldsymbol{k},\boldsymbol{k}')$, starting from a real-space representation, $g_{abc}(\boldsymbol{R},\boldsymbol{R}')$, is obtained with a tensor operation in the form
\begin{equation}
   f_{\alpha\beta\gamma}(\boldsymbol{k},\boldsymbol{k}')
   =
   \sum_{\boldsymbol{R} \boldsymbol{R}'} 
   \sum_{a b c} 
   e^{i\boldsymbol{k}\cdot\boldsymbol{R} + i\boldsymbol{k}' \cdot \boldsymbol{R}'} 
   U_{\alpha a}(\boldsymbol{k}') g_{abc}(\boldsymbol{R},\boldsymbol{R}') U_{\beta b}(\boldsymbol{k}) U_{\gamma c}(\boldsymbol{k}'-\boldsymbol{k}) \;,
\end{equation}
where $U$ are rectangular matrices indicating Wannier rotation matrices or phonon eigenvectors, $a$, $b$ and $c$ are indices labeling Wannier functions or atomic displacements, $\alpha$, $\beta$ and $\gamma$ are Bloch band indices, $\boldsymbol{k}$ is a wavevector, and $\boldsymbol{R}$ is a Bravais lattice vector.

In order to maximise performance, we first note that this expression typically appears inside a double loop on $\boldsymbol{k}$ and $\boldsymbol{k}'$.
Therefore, if $\boldsymbol{k}$ is the outer loop, it is convenient to precompute the quantity
\begin{equation}
   \tilde{f}_{\alpha b c}(\boldsymbol{k},\boldsymbol{R}')
   =
   \sum_{\boldsymbol{R} b} 
   e^{i\boldsymbol{k}\cdot\boldsymbol{R}} 
   g_{abc}(\boldsymbol{R},\boldsymbol{R}') U_{\beta b}(\boldsymbol{k}) \;,
\end{equation}
at fixed value of $\boldsymbol{k}$, and then complete the transformation inside the loop over $\boldsymbol{k}'$.
It is necessary to adopt the most convenient tensor contractions in order to minimize the size of loops needed to perform the interpolation.

Additionally, because GPUs are extremely effective at performing tensor operations, we implemented the calculation of the coupling with the Kokkos library \cite{osti_1106586}.
This library provides a hardware-agnostic programming model for parallelizing operations, so that tensor operations are implemented once following Kokkos' syntax and then, at compilation time, the code is optimized for the desired architecture.
The code can thus run both on standard CPU-only architecture as well as on GPUs when available without code duplication.
We therefore achieve strong parallel performance across a varied range of underlying architectures.

In our preliminary tests, we observed that the interpolation at fixed values of $\boldsymbol{k}$ and $\boldsymbol{k}'$ is extremely fast on a GPU, causing communication latency between GPU and CPU to be the dominant bottleneck.
To counteract this effect, we increase the workload on the GPU by performing the interpolation for a fixed value of $\boldsymbol{k}$ and a set of values for $\{ \boldsymbol{k}' \}$ at once.
The size of the set $\{ \boldsymbol{k}' \}$ is optimized at runtime so that most of the on-board GPU's memory is used.
Finally, as the size of the electron-phonon coupling matrix in the Wannier representation $g_{bb'\nu}(\boldsymbol{R}_e,\boldsymbol{R}_p)$ may be too large to fit in a single GPU's memory, we adopt an additional layer of MPI parallelization.
The electron-phonon coupling can be distributed over the index $\boldsymbol{R}_e$ across a ``pool'' of MPI processes, and we allow the user to specify the size of the pool at runtime through the command line.

The electron-phonon matrix interpolation of polar materials takes an additional correction, specified in Eq. \ref{eq:polarlong}, due to long-range interaction.
In order to optimize the evaluation of this long-range correction, we note that most terms, except the overlap matrix $\left\langle \psi_{\boldsymbol{k}+\boldsymbol{q},b'}\left| e^{i\left(\boldsymbol{q}+\boldsymbol{G}\right)\cdot\boldsymbol{r}} \right|\psi_{\boldsymbol{k}+\boldsymbol{q},b}\right\rangle$, depend exclusively on the phonon wavevector $\boldsymbol{q}$.
We therefore precompute these factors for all possible values of $\boldsymbol{q}$ once, at the start of the coupling interpolation.
Next, we note that the overlap term can be evaluated as $\left\langle \psi_{\boldsymbol{k}+\boldsymbol{q},b'}\left| e^{i\left(\boldsymbol{q}+\boldsymbol{G}\right)\cdot\boldsymbol{r}} \right|\psi_{\boldsymbol{k}+\boldsymbol{q},b}\right\rangle =
\big[ U(\boldsymbol{k}+\boldsymbol{q}) U^{\dagger}(\boldsymbol{k}) \big]_{b'b}$, and, since the $U$ matrices are readily available for the calculation of the short-range component of the electron-phonon matrix, it is readily possible to add this contribution without much overhead to the calculation.


\subsection{Quantum ESPRESSO interface for electron-phonon coupling}
\label{section:qePatch}
Phoebe interfaces with QE \verb+ph.x+ and Wannier90 in order to obtain the relevant ab-initio input for the electron-phonon coupling calculations.
However, this interface presented an important technical challenge.
As mentioned in Section \ref{Section:ElectronProperties}, gauge fixing is a critical component of the Wannier interpolation.
In order for the Wannier interpolation to work, it is critical that the gauge of the wavefunctions used by these two separate codes stays the same throughout the calculations.
Unfortunately, there is no built-in QE function to fix the gauge of a wavefunction and therefore the wavefunction may take a random phase every time it is computed.
As a result, we developed a gauge-fixing procedure that ensures the wavefunction has a consistent gauge across \verb+pw.x+ (the QE code that finds the ground state wavefunctions), \verb+ph.x+ (the QE code that computes the electron-phonon coupling) and \verb+wannier90+ (which computes the maximally-localized Wannier functions).

Therefore, in addition to the main Phoebe code, we provide a patched version of QE that must be used to compute electronic properties, which guarantees the gauge of the wavefunctions is appropriately taken into account.
In order to explain this gauge fixing procedure, we first recall that QE is a plane-wave ab-initio code, so that the wavefunction is computed as
\begin{equation}
    \psi_{\boldsymbol{k}b} (\boldsymbol{r})
    =
    \sum_{\boldsymbol{G}}
    c_{\boldsymbol{k}b}(\boldsymbol{G})
    e^{i\boldsymbol{G} \cdot \boldsymbol{r} + i\boldsymbol{k} \cdot \boldsymbol{r}} \;,
\end{equation}
where $\boldsymbol{G}$ runs over a set of reciprocal lattice vectors (typically defined up to a user-defined cutoff value on $|\boldsymbol{k}+\boldsymbol{G}|^2$), $c$ are the plane wave coefficients, and $\boldsymbol{r}$ is the position.
The subroutine \verb+c_bands()+ of QE is responsible for building and diagonalizing the Hamiltonian matrix, and therefore also computes the eigenvalues $\epsilon_{\boldsymbol{k}b}$ and the plane wave coefficients $c_{\boldsymbol{k}b}(\boldsymbol{G})$.
The phase arbitrariness manifests as a random phase on the complex plane-wave coefficients $c_{\boldsymbol{k}b}(\boldsymbol{G})$.
In our patch, we modified \verb+c_bands()+ so that if \verb+c_bands()+ is called by \verb+pw.x+ and QE has found the ground state wavefunction (i.e. a scf calculation), we save $c_{\boldsymbol{k}b}(\boldsymbol{G})$ to file.
These coefficients will serve as a reference wavefunction to fix the gauge of all subsequent calculations.
If \verb+c_bands()+ is called by another code, for example, when generating auxiliary data for Wannier90 or \verb+ph.x+, we read the reference plane-wave coefficients and use them to fix the gauge of the newly computed wavefunction.
In this way, we can ensure that - at the phase level - the same wavefunction is used throughout these multiple calculations.

An additional complexity in this patch is caused by the use of symmetries in the ab-initio code.
The reference plane-wave coefficients built in a \verb+pw.x+ calculation are typically computed only for wavevectors $\{ \boldsymbol{\tilde{k}} \}$ in the irreducible BZ wedge.
Therefore, we save to file only the coefficients $c_{\boldsymbol{\tilde{k}}b}(\boldsymbol{G})$ computed at irreducible points.
The wavefunction at a symmetry-equivalent wavevector $\boldsymbol{k}$ can be reconstructed from the irreducible wavevector $\boldsymbol{\tilde{k}} = R \boldsymbol{k}$ as
\begin{equation}
    \psi_{R\boldsymbol{k}b}(\boldsymbol{r})
    =
    \psi_{\boldsymbol{k}b}(R^{-1}(\boldsymbol{r}-\boldsymbol{t})) \;,
\end{equation}
which in terms of plane-wave coefficients implies
\begin{equation}
    c_{R\boldsymbol{k}b}(\boldsymbol{G})
    =
    e^{-i(R\boldsymbol{k}+\boldsymbol{G})\cdot t}
    c_{\boldsymbol{k}b}(R^{-1}\boldsymbol{G}) \;,
\end{equation}
where $S=\{R,t\}$ is a crystal symmetry as introduced in Section \ref{Section:Symmetries}.
An additional symmetry that is taken into account is the time reversal operation
\begin{equation}
    \psi_{-\boldsymbol{k}b}(\boldsymbol{r})
    =
    \psi^*_{\boldsymbol{k}b}(\boldsymbol{r}) \;,
\end{equation}
which in terms of plane-wave coefficients is
\begin{equation}
    c_{-\boldsymbol{k}b}(\boldsymbol{G})
    =
    c^*_{\boldsymbol{k}b}(-\boldsymbol{G}) \;.
\end{equation}
Another constraint comes from the translational symmetry of the crystal, i.e.
\begin{equation}
    \psi_{\boldsymbol{k}b}(\boldsymbol{r})
    =
    \psi_{\boldsymbol{k}+\boldsymbol{G}',b}(\boldsymbol{r}) \;,
\end{equation}
or equivalently
\begin{equation}
    c_{\boldsymbol{k}+\boldsymbol{G},b}(\boldsymbol{G})
    =
    c_{\boldsymbol{k}b}(\boldsymbol{G}+\boldsymbol{G}') \;.
\end{equation}
Capability for using symmetries in the presence of magnetism are currently being developed.
Using these relations and knowing the wavefunction at $\boldsymbol{\tilde{k}}$, it is possible to reconstruct the wavefunction at any symmetry-equivalent wavevector $\boldsymbol{k}$.

However, in QE, the wavefunction cannot be built from symmetries just by mapping $c_{\boldsymbol{\tilde{k}},b}(\boldsymbol{G})$ into $c_{\boldsymbol{k},b}(\boldsymbol{G})$.
In fact, the set of $\boldsymbol{G}$ vectors is chosen such that $|\boldsymbol{G}+\boldsymbol{k}|^2 < \delta$, where $\delta$ is a user-specified cutoff.
As a result, we cannot map all $c_{\boldsymbol{\tilde{k}},b}(\boldsymbol{G})$ into $c_{\boldsymbol{k},b}(\boldsymbol{G})$ as some plane-wave coefficients might be missing at the rotated wavevector, causing errors in the normalization of the wavefunction.
To address this, we first rotate the reference wavefunction $\ket{\psi^{\mathrm{ref}}_{\boldsymbol{\tilde{k}}b}}$ into the desired wavevector $\boldsymbol{k}$ finding $\ket{\psi^{\mathrm{ref}}_{\boldsymbol{k}b}} = \ket{S\psi^{\mathrm{ref}}_{\boldsymbol{\tilde{k}}b}}$.
Next, we compute the overlap matrix $O$ as
\begin{equation}
    O_{\boldsymbol{k} bb'} 
    =
    \langle \psi^{\mathrm{ref}}_{\boldsymbol{k},b} | \psi^{\mathrm{arb-gauge}}_{\boldsymbol{k},b'} \rangle_{bb'} \;,
\end{equation}
where $\ket{\psi^{\mathrm{arb-gauge}}_{\boldsymbol{k},b} }$ is the wavefunction with an arbitrary gauge that has been computed by QE at wavevector $\boldsymbol{k}$.
With a complete basis set $O$ would be a unitary matrix, however, due to the finite set of $\boldsymbol{G}$-vectors, the unitary property is violated by a small amount $\Delta$ defined as
\begin{equation}
    \Delta = 1 - OO^{\dagger} \;.
\end{equation}
We can re-enforce the unitary property of $O$ by first noting that $O$ is expected to be semi-positive definite (due to the lack of completeness of the finite set of $\boldsymbol{G}$ vectors).
We then write
\begin{align}
    1 
    &= OO^{\dagger} + \Delta  \\
    &= 
    OO^{\dagger} + (OO^{\dagger}+\Delta) \Delta (OO^{\dagger}+\Delta) \\
    &\approx 
    OO^{\dagger} + OO^{\dagger} \Delta OO^{\dagger} \\
    &= 
    OLL^{\dagger}O^{\dagger}
\end{align}
where $L$ is the Cholesky decomposition of the matrix $LL^{\dagger}=1+O^{\dagger}\Delta O$ and introduced an approximation that has an error of order $\Delta^3$ (expected to be small for well-converged calculations).
By construction, the matrix $OL$ is a unitary matrix which can now be used to rotate the wavefunction computed at a symmetry-equivalent point to the desired gauge, while preserving the orthonormality of the wavefunctions. 
We therefore fix the gauge of the wavefunction at $\boldsymbol{k}$ by applying the rotation
\begin{equation}
    \psi^{\mathrm{fixed-gauge}}_{\boldsymbol{k}} = OL \psi^{\mathrm{arb-gauge}}_{\boldsymbol{k}} \;.
\end{equation}

This procedure makes the wavefunctions obey the symmetries of the crystals and greatly simplifies the electron-phonon coupling calculation.
As explained in Ref. \cite{PhysRevB.76.165108}, if the wavefunction obeys crystal symmetries, the electron-phonon coupling satisfies the symmetry relation
\begin{equation}
    g_{bb'\nu} (S\boldsymbol{k},\boldsymbol{q})
    =
    g_{bb'\nu} (\boldsymbol{k},S^{-1}\boldsymbol{q}) \;,
\end{equation}
where $S$ is a symmetry operation of a crystal and
\begin{equation}
    g_{bb'\nu} (\boldsymbol{k},\boldsymbol{q})
    =
    g_{bb'\nu} (\boldsymbol{k}+\boldsymbol{G},\boldsymbol{q}+\boldsymbol{G}) \;,
\end{equation}
due to the periodicity of the crystal.
As a result, we can compute the electron-phonon coupling on a set of irreducible wavevectors, then construct the remaining matrix elements by means of the symmetry relations, reducing the burden of the ab-initio calculation.

We stress that this QE patch avoids the need for re-implementing calculations of the electron-phonon matrix elements $\langle \psi_{\boldsymbol{k}+\boldsymbol{q},b'} | \partial_{\boldsymbol{q}\nu} V | \psi_{\boldsymbol{k}b} \rangle$, which are already computed in the code \verb+ph.x+.
As a result, Phoebe can readily take advantage of all features supported by \verb+ph.x+, such as ultra-soft or PAW pseudopotentials \cite{ESPRESSO2009}, U corrections \cite{QE-2017} and others, without the need for re-implementing them.
The only modification necessary in the code \verb+ph.x+ has been the introduction of a subroutine that unfolds the symmetry of the coupling $g$ and writes it to file output.
The Wannier90 code does not need to be modified, as it simply reads the wavefunctions pre-computed by QE.
As a last detail, we stress that the gauge problem must also be taken care when dealing with the phonon eigenvector $z$, and the eigenvector $z$ present in Eq. \ref{blochToWannierG} must be exactly the same phonon eigenvector (phase factor included) used by \verb+ph.x+ to compute the electron-phonon coupling.
Additionally, $z$ must obey the symmetries of the crystal, as discussed in Ref. \cite{PhysRevB.76.165108}.

\subsection{Workflow Overview}

\begin{figure}[H]
    \centering
    \includegraphics{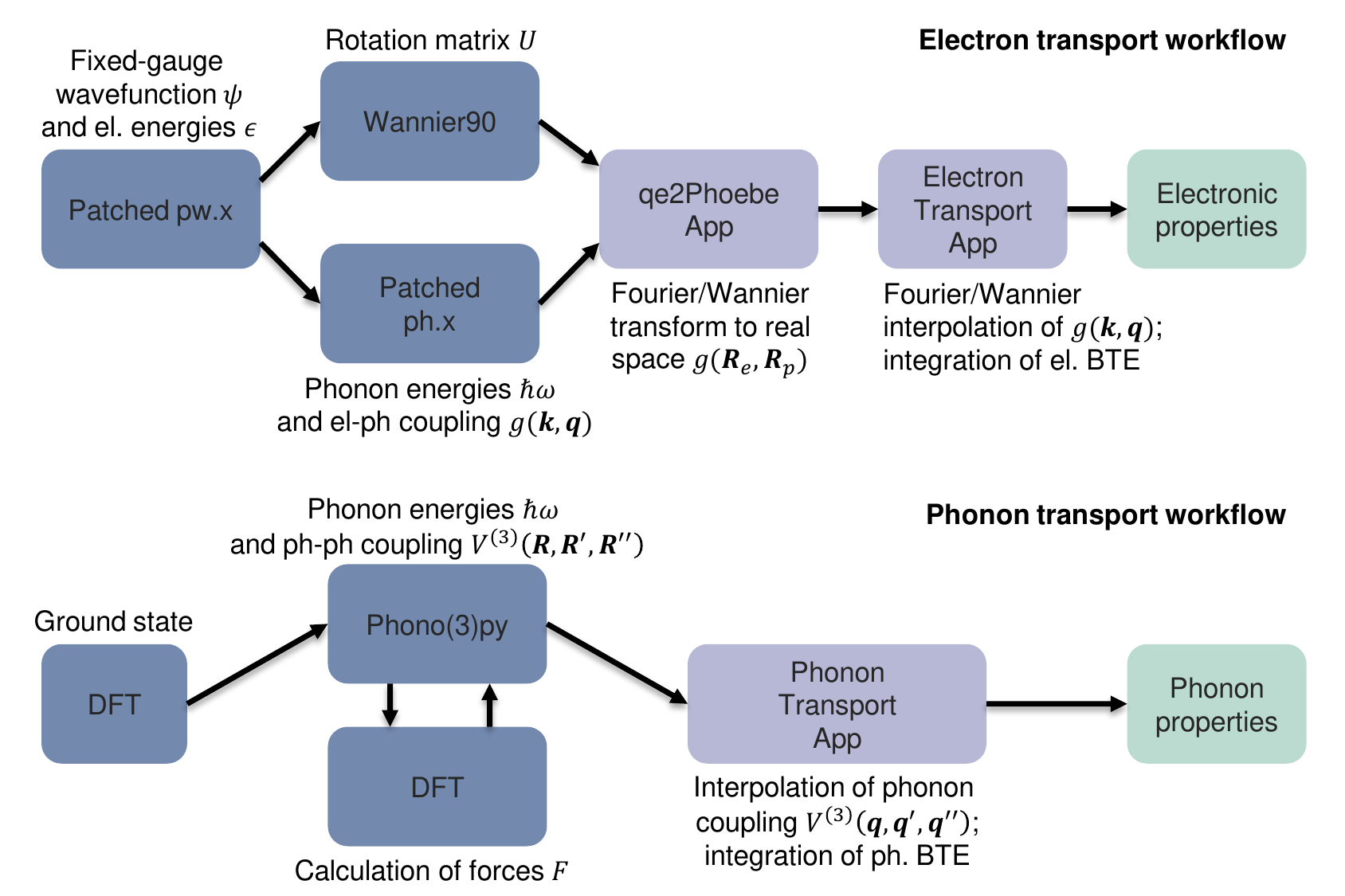}
    \caption{Schematic depiction of the user workflows to compute electronic transport properties with Phoebe (top panel) or phonon transport properties (bottom panel). Generally, the user must first perform ab-initio simulations (blue boxes), using our patched version of Quantum ESPRESSO for studying electron-phonon interactions or any DFT code that can provide anharmonic phonon interactions. Next, shown in purple boxes, the user must launch the Phoebe executable. The electronic transport properties calculation must be preceded by a one-time post-processing of the electron-phonon coupling taken in output from Quantum ESPRESSO, while the phonon transport calculation can be directly started from the output of ab-initio calculations. Finally, results are provided in output either in the standard output file, or as a parsable JSON format.}
    \label{fig:workflow}
\end{figure}

Having introduced all the major features and computational structures of Phoebe, we demonstrate how the code can be used to calculate transport properties of materials. In Fig. \ref{fig:workflow} we schematically show a user's workflow for computing electronic and thermal transport properties with Phoebe.
The electronic transport workflow, as shown in the top panel, starts with the user going through the ab-initio calculation.
The first step requires the calculation of the ground state wavefunction and electronic energies.
Due to the wavefunction gauge requirements explained in Section \ref{section:qePatch}, the ground state calculation must be performed using our patched version of QE.
Next, the user proceeds to compute the phonon properties with the patched version of \verb+ph.x+.
The output of this calculation, specifically files containing the force constants and the electron-phonon coupling, are later passed to Phoebe.
At the same time, the user must also compute the Wannier functions with Wannier90, which outputs  the real-space representation of the electronic Hamiltonian and the $U$ rotation matrices that must be used to rotate the wavefunction to the Wannier gauge.
After these ab-initio calculations have concluded, the user must run the \verb+qe2PhoebeApp+, which uses the data from the ab-initio calculations to compute the electron-phonon coupling in the Wannier representation $g(\boldsymbol{R}_e,\boldsymbol{R}_p)$ starting from $g(\boldsymbol{k},\boldsymbol{q})$ computed by \verb+ph.x+, as described in Eq. \ref{blochToWannierG}.
This operation needs to be run only once per material, and generates an output (HDF5) file with the electron-phonon coupling in real space, ready to be used for the interpolation.
Finally, the user has all the input data necessary to compute the electron transport properties by launching the \verb+electronWannierTransportApp+.
Results are output as JSON files, easily parsed and analyzed by the user.

The workflow for computing the phonon transport properties is shown in the lower panel of Fig. \ref{fig:workflow}.
First, the user must select an ab-initio code (from any that work with phono3py or ShengBTE) to calculate the equilibrium ground state of the crystal.
After a reference equilibrium crystal structure is found, the user runs an external library, e.g. phonopy and phono3py or ShengBTE, in conjunction with the ab-initio code to generate all atomic displacements needed to compute the second and third derivatives of the total energy (see Sec. \ref{phononProperties}).
Once both $V^{(2)}$ and $V^{(3)}$ are computed, the \verb+PhononTransportApp+ is ready to be launched.
This app will interpolate the phonon-phonon coupling, compute phonon lifetimes, and then solve the BTE to evaluate the phonon transport properties such as the thermal conductivity.
As in the electronic transport case, the results are written in easily parsable JSON files that can be conveniently analyzed.

\section{Results}

Next, we showcase the calculation of transport properties by applying these two computational workflows to GaN, a semiconductor often used for high-power applications.



\subsection{GaN}
Gallium nitride (GaN) is a well-studied semiconductor that crystallizes in a wurtzite structure.
We calculate the electronic structure of this material via QE using optimized norm-conserving Vanderbilt pseudopotentials \cite{hamann2013optimized, van2018pseudodojo} and the local-density approximation exchange-correlation functional \cite{perdew1992accurate}.
For the phonon workflow, we construct the force constants in a 4$\times$4$\times$3 supercell, and the third derivative using a 3$\times$3$\times$2 supercell.
The supercell DFT calculation is performed using a 2$\times$2$\times$2 mesh for the $\boldsymbol{k}$-point integration and a wavefunction cutoff of 110 Ry.
Next, the phonon BTE is built using the adaptive smearing method for approximating the Dirac-delta and converged with a 15$\times$15$\times$15 mesh of $\boldsymbol{q}$-points.
For the electron transport workflow, ab-initio properties are computed using a 110 Ry plane-wave energy cutoff, a 12$\times$12$\times$7 coarse mesh for both $\boldsymbol{k}$- and $\boldsymbol{q}$-points and the same pseudopotential choice. 
The maximally-localized Wannier function basis for this material was established by starting with an initial guess of 14 random Wannier centers. 
The electronic BTE is build using an adaptive smearing scheme (Sec. \ref{dirac_delta_section}), and states are discarded with the population window described in ~\ref{Section:BandStructure}.
The BTE is integrated using a variable mesh of k-points, starting from the densest mesh of 170$\times$170$\times$106 at the lowest temperature of 100K and gradually decreased down to 90$\times$90$\times$56 at the highest temperature of 450K, as the convergence properties depend on temperature. 

\begin{figure}
    \centering
    \includegraphics[width=0.8\textwidth]{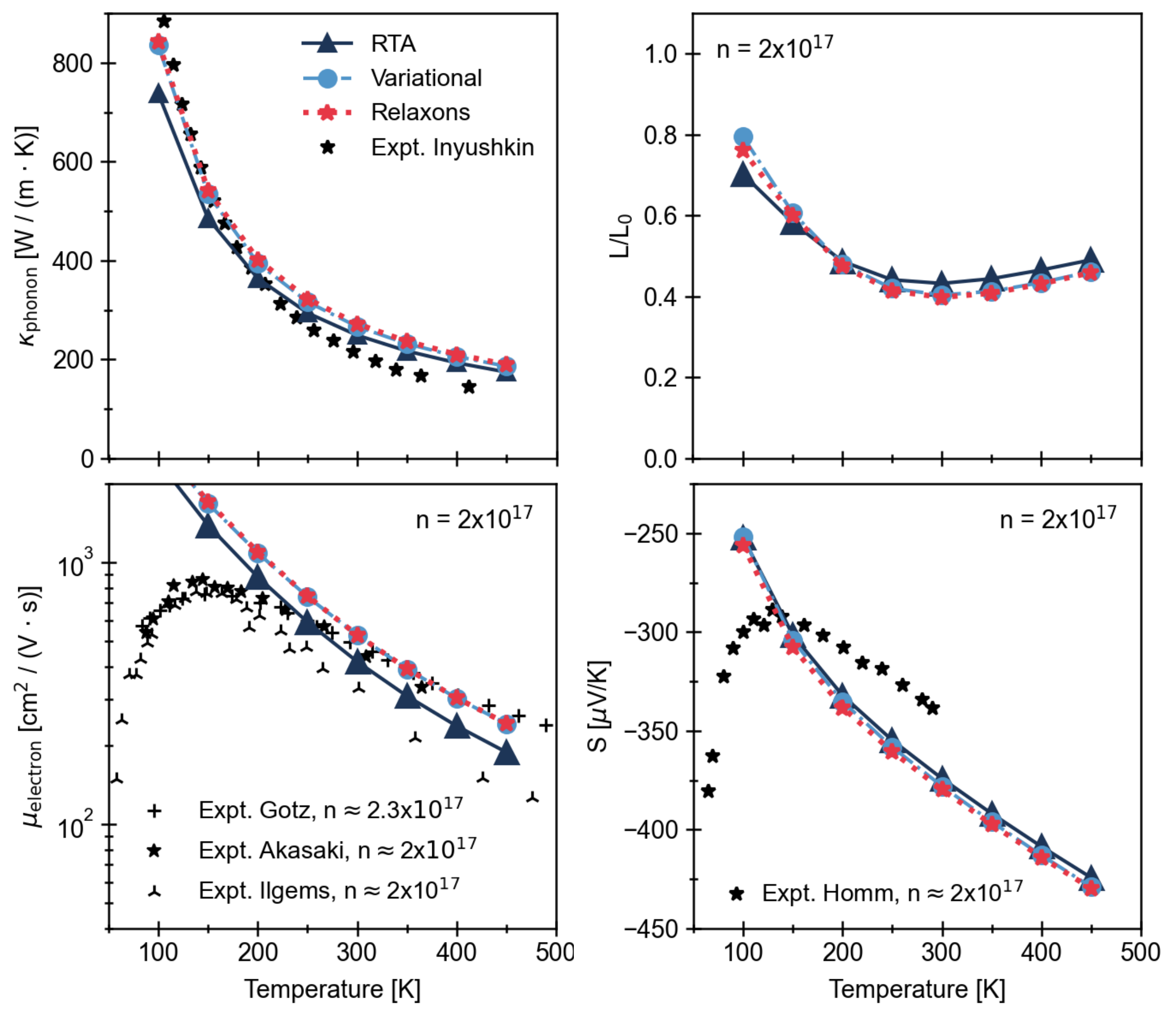}
    \caption{The calculated lattice thermal conductivity (top left), electron mobility (bottom left), Lorenz number (top right), and Seebeck coefficient (bottom right) (with $\mu$, $L$, and $S$ calculated for carrier density n=2x$10^{17} cm^{-3}$) of GaN along the a-direction crystal axis determined using the RTA, variational and relaxons solutions to the BTE as implemented in Phoebe. Experimental data for $\kappa_{phonon}$ \cite{inyushkin2020GaNKappa}, $\mu_{electron}$ \cite{gotz1996activation,akasaki1989effects,ilegems1973electrical}, and S \cite{homm2008seebeck} are shown for comparison.}
    \label{fig:GaN_v_T}
\end{figure}    

In Fig. \ref{fig:GaN_v_T} we plot some of the transport properties of GaN as a function of temperature.
In the top left panel of Fig. \ref{fig:GaN_v_T}, we plot the lattice thermal conductivity as a function of temperature, and compare results of the BTE at various levels of approximation with the experimental values \cite{inyushkin2020GaNKappa}.
We find a good agreement between our simulations and experiments, validating the results of this work.
The RTA solution tends to underestimate the thermal conductivity predicted by exact BTE, as expected due to the variational principle underlying the formalism \cite{ziman, fugalloVariational}.
To solve the phonon BTE exactly, we use the variational and the relaxons solvers, mainly to note how the two solvers are able to provide, as expected, approximately the same solution to the BTE.

The electron drift mobility at a n-carrier concentration of about $2\cdot 10^{17}$ cm$^{-3}$ is plotted versus temperature in the bottom left panel of Fig. \ref{fig:GaN_v_T} and compared against some of the available experiments \cite{gotz1996activation, akasaki1989effects, ilegems1973electrical}.
Here we distinguish two regimes.
At higher temperatures, where a sufficient number of phonons is thermally excited, we expect electron-phonon scattering to be the dominant factor limiting charge transport in GaN.
Indeed, for temperatures above approximately 150K, we obtain a good agreement with experimental values, and correctly reproduce the decreasing trend of mobility with respect to temperature.
At lower temperatures we observe a departure in computed mobility from experimental values, which tend to decrease as temperature goes to zero, while our results are showing the opposite trend.
This is likely due to the fact that at lower temperature fewer phonons are thermally excited, and electrons have a higher probability of scattering against defects or impurities in the crystal, which have not been taken into account for this simulation.
We also note that the RTA already provides a good approximation to the mobility (and also for the Seebeck coefficient and the resistivity), and the variational and relaxon solvers provide essentially the same estimates.
Overall, we observe a good agreement with experiments for simulations in the phonon-limited regime of charge transport.

Additionally, in the top right panel of Fig. \ref{fig:GaN_v_T}, we include a calculation of the Lorenz ratio, defined as $L/L_0 = \frac{k_e}{\sigma T L_0}$, where $k_e$ is the electronic contribution to the thermal conductivity, and $L_0$ is the standard Lorenz constant for an ideal metal [GIVE VALUE].
In GaN we confirm that $L/L_0$ shows sensible behavior and is close to unity, i.e. the limit in which the Wiedemann-Franz law is obeyed.
A deviation from $L/L_0 = 1$ is to be expected, since the Wiedemann-Franz law is typically more closely followed by metals.

In Fig. \ref{fig:GaN_v_T}, bottom right panel, we plot the Seebeck coefficient as a function of temperature.
Here, we obtain a good agreement between our results and the experimental values for large temperatures \cite{homm2008seebeck}, i.e. larger than 150K.
Since the Seebeck coefficient is strongly related to the density of states of the material \cite{PhysRevB.79.153101}, the results seem to indicate that the DFT simulation provided a good approximation to the electronic band structure.
Also in this case, we observe a departure between our results and experiments at the lowest temperatures.
Electronic scattering by defects is not expected to change the Seebeck coefficient significantly \cite{PhysRevB.101.075202}.
Instead, this deviation could be attributed to phonon drag, which for example has been investigated in GaAs \cite{PhysRevB.101.075202}, and often manifests with a sudden change in the Seebeck coefficient at low temperatures.


\begin{figure}
    \centering
    \includegraphics[width=1.0\textwidth]{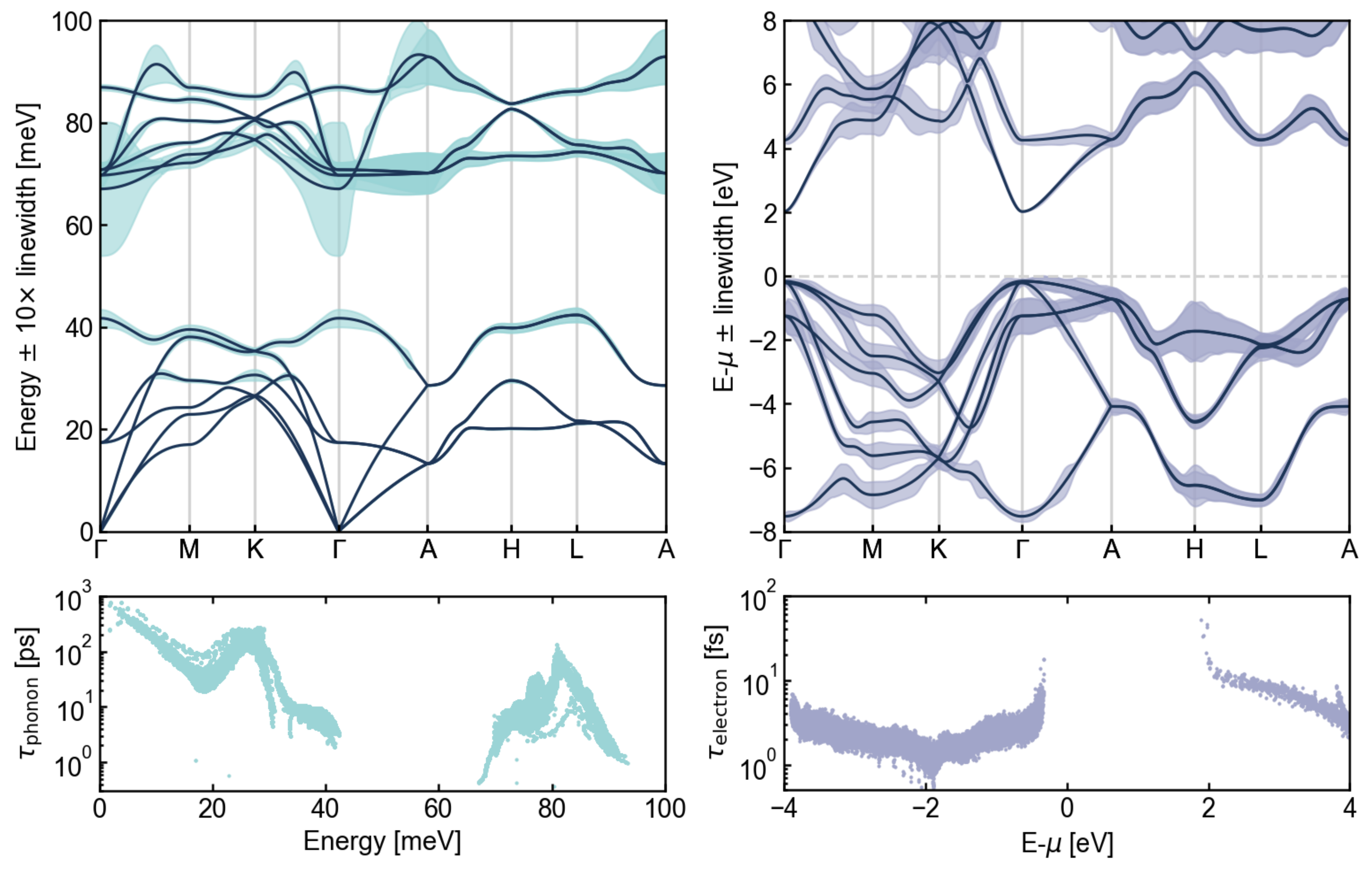}
    \caption{Plots of phonon (left panels) and electron (right panels) linewidths and lifetimes of GaN. In the top panels, linewidths are plotted as the thickness of lines against the band structure of the carrier on a high-symmetry path of the Brillouin zone. In the bottom panels, we plot lifetimes as a function of carrier energy. The reference zero energy corresponds to the chemical potential.}
    \label{fig:GaN_tau}
\end{figure}

In the top panels of Fig. \ref{fig:GaN_tau} we plot the phonon (left panels) and electronic (right panels) band structure.
On top of the band structure, we superimpose the carrier linewidths computed at a temperature of 300K. 
These pictures are produced with \verb+Python+ scripts that have been included in the Phoebe release for user's convenience.
The top panel provides an easy way to identify some particular bands or areas of the Brillouin zone which tend to scatter the most, as the linewidths are a measure of the total scattering rate.
In the upper-left panel of \ref{fig:GaN_tau}, we can observe how optical phonons tend to scatter more, especially at the $\Gamma$ or $A$ high-symmetry points.
In the upper-right panel of Fig. \ref{fig:GaN_tau} instead we can observe how the holes close to the chemical potential have larger linewidths than electrons close to the band gap, but otherwise we don't observe any pocket of large scattering rates.
The bottom panels describe the energy-dependence of the lifetimes, which sometimes is useful to derive simplified models of transport.
For phonons, in the lower-left panel of Fig. \ref{fig:GaN_tau}, we can clearly see that lifetimes are largest for the three acoustic phonon modes whose energies tend to zero at the $\Gamma$ point.
The calculated phonon lifetimes tend to decrease as the energy increases.
However, one can notice how the energy dependence cannot be described with a simple empirical power-law expression: an accurate microscopic description, as done here, is often necessary to describe phonon lifetimes.
In the lower-right panel \ref{fig:GaN_tau}, we instead show the energy dependence of electron lifetimes due to electron-phonon scattering.
We observe a spike in the electron lifetimes for carriers closest to the band edge due to the vanishing density of states; often these carriers are responsible for determining the transport properties of a material. 
Away from the gap, the relaxation times simply show a weaker energy dependence.


\subsection{Performance and Scaling}

\begin{figure}
    \centering
    \includegraphics{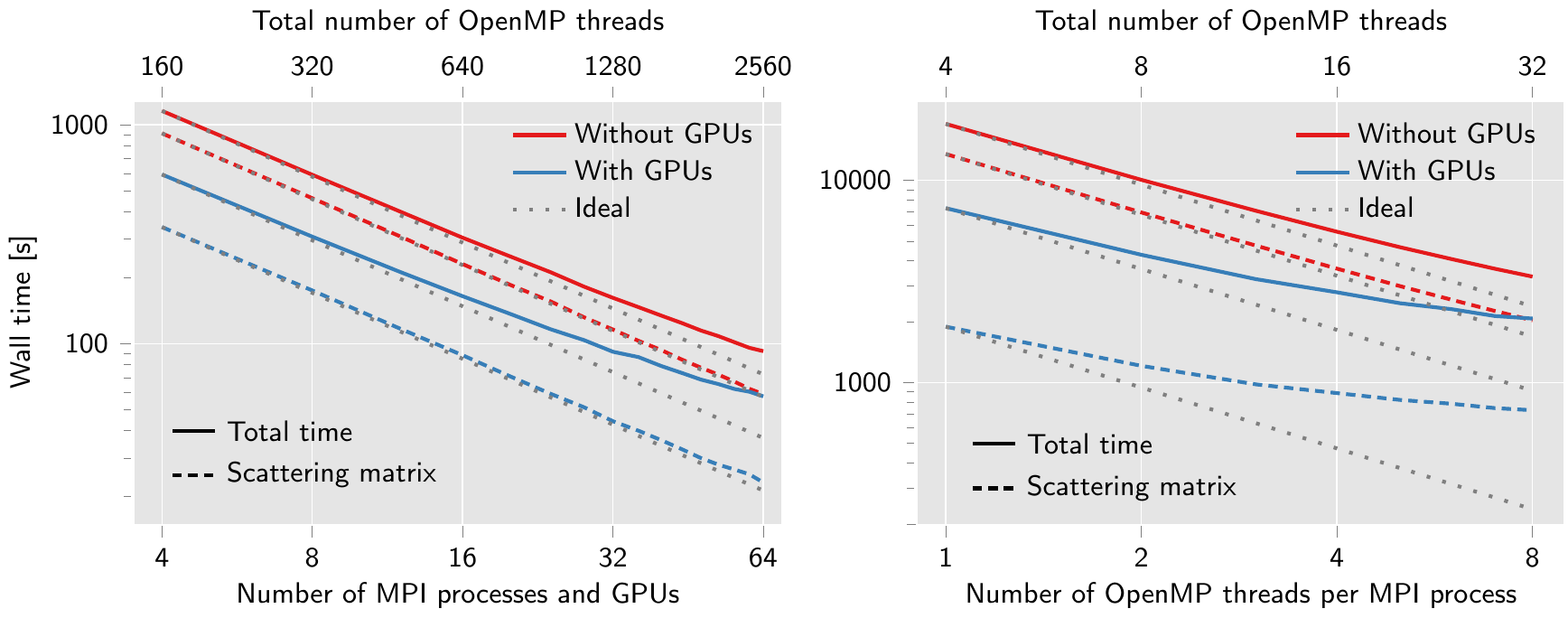}
    \caption{Strong scaling with MPI and OpenMP. The solid lines show the total execution time of Phoebe for the calculation of electronic transport properties of GaN, while the dashed lines show the time used to calculate the scattering matrix. 
    On the left panel, the number of MPI processes is varied while the number of OpenMP threads is kept constant, and vice-versa on the right panel. 
    In blue lines, we show results when the GPU is used (one V100 GPU per MPI process), which is contrasted with results in red lines performed on the same computational nodes but without GPUs. GPUs speedup simulations by approximately a factor of three.}\label{fig:scaling}
\end{figure}

Phoebe uses MPI, OpenMP and Kokkos for scaling and acceleration at different levels. 
MPI enables large-scale simulations on multiple nodes, while OpenMP allows shared-memory parallelization within one node. 
As explained in Section \ref{Section:Kokkos}, the interpolation of the particle interaction is implemented with Kokkos, so that if no GPU is available, Kokkos's OpenMP backend will add to the intra-node parallelization, otherwise its CUDA, HIP, SYCL and OpenMPTarget backends compile the code for GPUs from different vendors.

In this section, we show some examples of the computational performance of Phoebe.
As a benchmark, we take the GaN calculation of electronic transport properties discussed in the previous section, and integrate the BTE at a temperature of 300~K on a grid of \(90\times90\times56\) wavevectors.
For this test, we reduced the coarse ab-initio grid of wavevectors to \(7\times7\times4\).
The simulation is run on the TACC Longhorn supercomputer, where each node has 4 NVIDIA V100 GPUs and 2 IBM Power 9 AC922 20-core CPUs.

In Fig.~\ref{fig:scaling} we plot the scaling of the \verb+ElectronWannierTransportApp+ with respect to the two levels of parallelization (MPI and OpenMP), while also demonstrating how the GPU accelerates the calculation (with 1 GPU per MPI process).
The solid lines show the total wall time of the calculation as a function of the number of MPI processes (left panel) or OpenMP threads per MPI thread (right panel), while dashed lines show the wall time for the calculation of the scattering matrix (typically the most computationally intensive algorithm present in the code).
Dotted lines represent the ideal scaling of a calculation -- a straight line in this logarithmic scale -- and the color code labels simulations performed with (blue) or without (red) GPU acceleration.

In the left panel of Fig.~\ref{fig:scaling}, we test the MPI scaling by increasing the number of nodes used, with 4 MPI ranks, 4 GPUs and 160 OpenMP threads per node.
When running without GPUs, the overall scaling of the simulation is very close to ideal.
In this case, the calculation of the scattering matrix dominates the total cost of the simulation and since it's efficiently distributed across MPI processes over the wavevectors, we achieve a good scaling performance.
The GPUs, shown in blue lines, greatly accelerate the calculation of the scattering matrix, which is now a much smaller contribution to the overall computation time.
Remarkably, we achieve a GPU acceleration of a factor 2 or 3 to the overall simulation wall time.
Dashed lines in Fig.~\ref{fig:scaling} show in detail the time taken by the construction of the scattering matrix.
This part, dominant for larger systems and grids, is very close to ideal scaling.

The right panel of Fig.~\ref{fig:scaling} showcases the scaling performance of Phoebe with respect to the number of OpenMP threads per MPI process.
Here, we used a single node with 4 GPUs and 4 MPI ranks, while the number of OpenMP threads per MPI rank is gradually increased from 1 to 8.
Because not all steps of the calculation can be parallelized with OpenMP (in particular, the input files reading at the start of the calculation), the total simulation time deviates from ideal scaling as more threads are used.
Nevertheless, we achieve good OpenMP scaling performance when GPUs are not used, especially for the construction of the scattering matrix.
When GPUs are used, we observe a significant deviation from the linear scaling, which can be traced to the lack of OpenMP scaling of the scattering matrix construction.
This is not surprising: since the majority of the scattering matrix calculation is offloaded to the GPU, it becomes largely independent of the number of OpenMP threads being used on the CPU, which are only responsible for minor precomputations and data transfer.


\begin{figure}
    \centering
    \includegraphics[width=0.5\textwidth]{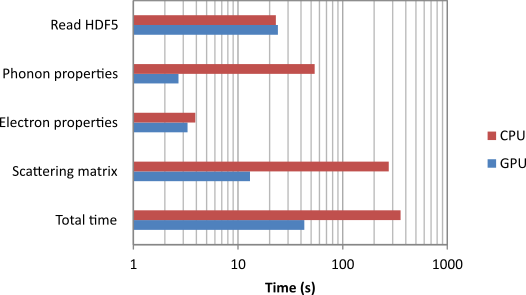}
    \caption{Performance breakdown during a simulation of GaN electronic transport properties, with and without GPUs. The simulation is run on a single node with 4 MPI ranks, 8 OpenMP threads and 4 GPUs. The first step is to read the HDF5 file, which only depends on file system speed. Phoebe then computes the energies, eigenvectors and velocities for the phonon and electron Hamiltonians at different wavevectors, and finally the scattering matrix. Overall, we find the GPU greatly accelerates the total computation time.}
    \label{fig:perfcomp}
\end{figure}

Figure~\ref{fig:perfcomp} shows the performance breakdown of the most significant steps in a single electronic transport calculation. 
This simulation is computed using a \(80\times80\times50\) \(k\)-mesh, and has been run on one node of Harvard University's Cannon supercomputer with 4 MPI processes and 8 OpenMP threads, where each node has 4 V100 GPUs and 2 Intel Xeon Gold 6142 16-core CPUs.
The simulation is profiled using the KokkosP Profiling tools.
A vast majority of the initial simulation time is spent reading input files, in particular the large HDF5 files containing the electron-phonon coupling in real space precomputed from Eq.~\eqref{blochToWannierG}. 
While file-reading cannot be accelerated with OpenMP or GPUs, we use parallel HDF5 to maximize the I/O speed.
In the next step of the calculation, Phoebe precomputes electron and phonon properties as described in sections~\ref{phononProperties} and~\ref{Section:ElectronProperties}, i.e. energies, eigenvectors, and velocities at all points \(\boldsymbol{k}\), \(\boldsymbol{q}\) and \(\boldsymbol{k}+\boldsymbol{q}\). 
This operation is trivially parallelized with both MPI processes and either OpenMP threads or a GPU, thus driving the overall scaling towards linearity. The GPU is particularly efficient for diagonalizing the phonon and electron Hamiltonians at different wavevectors by using the \texttt{cusolverDnZheevjBatched} function from the cuSOLVER library, which diagonalizes a batch of small matrices.
At this stage, we also precompute all factors in the long-range polar correction (Eq.~\ref{eq:polarlong}), except for the wavefunction overlap which is added during the calculation of the scattering rates. 
In this way, part of the calculation of the polar correction has a linear scaling with $\boldsymbol{q}$, rather than being embedded in a loop over $\boldsymbol{k}$ during the construction of the scattering matrix.
In the last part of the simulation, Phoebe computes the scattering matrix, which is the largest contribution to the total wall time.
When the GPU is being used, we can see how the calculation of the scattering matrix is significantly accelerated, to the point where the GPU calculation is about ten times faster than the calculation on CPU, and the GPU simulation is dominated by the cost of I/O.
The use of non-blocking MPI reductions (\texttt{MPI\_Ireduce}) minimizes the time spent on the CPU and further enhances the scaling.
The calculation of transport properties after the scattering matrix has been built is done at a negligible cost compared to the operations shown in Fig.~\ref{fig:perfcomp}.

Overall, we demonstrate that Phoebe successfully leverages multi-level parallelism and GPU acceleration even for a moderately expensive calculation. 
With just a single GPU node, the calculation is fast enough that it becomes dominated by unavoidable overhead tasks like file reading and data transfer.


\section{Conclusion}
In this article we present Phoebe, a high-performance software for characterising electron and phonon transport properties from first-principles.
It provides various tools to predict transport properties at various levels of theory and accuracy, such as the Boltzmann transport equation or models based on the Wigner distribution.
Phoebe interfaces with Phono3py and ShengBTE to obtain the first-principles coupling strength of the phonon-phonon interaction, and with QE for the electron-phonon interaction.
The accurate integration of the Boltzmann transport equation and related transport properties is achieved with Fourier/Wannier based interpolation techniques or with energy-averaging approximations.
Lastly, the code is optimized for speed and scaling, using a mix of parallelization levels based on MPI, OpenMP and GPU (through Kokkos).
The code has been made publicly available on Github (https://mir-group.github.io/phoebe/) under an open-source basis (MIT license).

The ongoing goal for this project is to provide the community with a scalable software to efficiently perform simulation of materials transport properties, while providing an extensible platform for implementing new methods as they are being developed.





\subsubsection*{Acknowledgments}
We acknowledge funding support from the Star-Friedman Fund for Promising Scientific Research, the Harvard Quantum Initiative, the Harvard Climate Change Solutions Fund, the STC Center for Integrated Quantum Materials, NSF Grant No. DMR-1231319. Computational resources were provided by the Harvard FAS Research Computing and TACC, allocation DMR20013. J.C. acknowledges support from the Department of Energy Computational Science Graduate Fellowship (DOE CSGF) under Award Number DE-FG02-97ER25308. A.J. acknowledges support from the Aker Scholarship. N.S.F. acknowledges support from Swiss National Science Foundation (project number P2EZP2\_178532).

\end{document}